%% file: ms.tex
\documentclass[numberedappendix]{emulateapj}
\usepackage{amsmath,bm}
\usepackage{epstopdf}

\begin{document}

\title{Type Ia Supernova Light Curve Inference: \\ Hierarchical Models in the Optical and Near Infrared}
\author{Kaisey S. Mandel\altaffilmark{1}, Gautham Narayan, Robert P. Kirshner}
\affil{Harvard-Smithsonian Center for Astrophysics, 60 Garden St., Cambridge, MA 02138}
\altaffiltext{1}{kmandel@cfa.harvard.edu}
\slugcomment{Accepted for publication in ApJ, in press}

\newcommand{\dmB}{\Delta m_{15}(B)}
\newcommand{\dm}{\Delta m_{15}}
\newcommand{\Bmax}{B_{\text{max}}}   
\newcommand{\tBmax}{T_{B\text{max}}}
\newcommand{\var}{\text{Var}}
\newcommand{\cov}{\text{Cov}}
\newcommand{\nsn}{{N_\text{SN}}}
   
\begin{abstract}
We have constructed a comprehensive statistical model for Type Ia supernova (SN Ia) light curves spanning optical through near infrared (NIR) data. A hierarchical framework coherently models multiple random and uncertain effects, including intrinsic supernova light curve covariances, dust extinction and reddening, and distances.    An improved \textsc{BayeSN} MCMC code computes probabilistic inferences for the hierarchical model by sampling the global probability density of parameters describing individual supernovae and the population.   We have applied this hierarchical model to optical and NIR data of 127 SN Ia from PAIRITEL, CfA3, CSP, and the literature.   We find an apparent population correlation between the host galaxy extinction $A_V$ and the the ratio of total-to-selective dust absorption $R_V$.   For SN with low dust extinction, $A_V \lesssim0.4$, we find $R_V \approx 2.5-2.9$, while at high extinctions, $A_V \gtrsim 1$, low values of $R_V< 2$ are favored.    The NIR luminosities are  excellent standard candles and are less sensitive to dust extinction.  They exhibit low correlation with optical peak luminosities, and thus provide independent information on distances. The combination of NIR and optical data constrains the dust extinction and improves the predictive precision of individual SN Ia distances by about $60\%$.
Using cross-validation, we estimate an rms distance modulus prediction error of 0.11 mag for SN with optical and NIR data versus 0.15 mag for SN with optical data alone.  
Continued study of SN Ia in the NIR is important for improving their utility as precise and accurate cosmological distance indicators.
\end{abstract}
\keywords{distance scale Ð- supernovae: general --  methods: statistical}

\section{Introduction}

Type Ia supernova (SN Ia) rest-frame optical light curves have been of great utility for measuring fundamental quantities of the 
universe.  As standardizable candles, they were critical to the detection of  cosmic acceleration \citep{riess98,perlmutter99}.    SN Ia  
have been used to constrain the equation-of-state parameter $w$ of dark energy  \citep{garnavich98b}, and recent efforts have 
measured $w$ to $\sim 10\%$, \citep{wood-vasey07, astier06, kowalski08, hicken09b, kessler09, freedman09, amanullah10}.   SN Ia have also been used to establish the extragalactic distance scale and measure the Hubble constant \citep{freedman01,jha99,riess05,riess09a,riess09b}, 

SN Ia distance indicators exploit 
empirical relations between peak optical luminosities of SN Ia and distance-independent measures such as light curve shape observed in the sample of nearby low-$z$ SN Ia \citep{hamuy96_29sne, riess99, jha06, hicken09a}.  Methods 
include $\Delta m_{15}(B)$ \citep{phillips93,hamuy96, phillips99, prieto06}, MLCS \citep{riess96, riess98, jha07}, ``stretch'' 
\citep{goldhaber01}, CMAGIC \citep{wang03}, SALT \citep{guy05, guy07}, and SiFTO \citep{conley08}.  One of the largest systematic uncertainties limiting the 
precision of distance estimates from rest-frame optical light curves is dust extinction in the host galaxy and the confounding of dust reddening with the intrinsic color variations of SN Ia \citep{conley07}.  Current approaches  differ conceptually and practically on how apparent colors, intrinsic colors, and dust effects are modeled.  While most methods make use of the optical luminosity-light curve width correlation, some methods, such as MLCS \citep{riess96, riess98, jha07}, attempt to separately model the intrinsic colors of the SN Ia and host galaxy dust reddening and extinction, whereas others model both effects with a single factor \citep[e.g. SALT2,][]{guy07}.

Early observations of SN Ia in the infrared were made by \citet{kirshner73, elias81, elias85, frogel87} and \citet{graham88}.  Observations of nearby SN Ia 
in the NIR  revealed that the peak near-infrared luminosities of SN Ia have a dispersion smaller than $0.20$ mag  
\citep{elias85, meikle00, krisciunas04a,krisciunas04c}.  
\citet{wood-vasey08} (hereafter WV08) compiled a sample of NIR SN Ia observations taken with the Peters 
Automated InfraRed Imaging TELescope \citep[PAIRITEL;][]{bloom06b}.  They found that the $H$-band peak absolute magnitude, had small scatter $\sigma(M_H) \approx 0.15$ mag, and could provide distance estimates competitive with those derived from optical light curve shapes.  The effect of dust extinction is significantly diminished at NIR wavelengths, relative to the optical.  The combination of optical and NIR observations of SN Ia light curves could lead to even better estimates of SN Ia distances \citep{krisciunas07}.

A significant source of puzzlement in the analysis of SN Ia light curves is the nature of the apparent color and brightness variations among supernovae, which are comprised of color and luminosity variations intrinsic to the SN Ia population, and also random reddening and extinction by dust in the host galaxies of SN Ia.   The function of dust absorption over wavelength is typically parameterized by the ratio of total to selective extinction, $R_V = A_V/(A_B - A_V)$.  This ratio has an average value of 3.1 for interstellar dust in the Milky Way Galaxy, although it can vary between 2.1 and 5.8 \citep{draine03}.   Studies of external galaxies have found similar extinction curves; for example, \citet{finkelman08, finkelman10} found average values of $R_V \approx 2.8$.
Early studies of SN Ia found values of $R_V < 1$ \citep[c.f.][for a review]{branch92}, although these analyses did not take into account relationships between the luminosity, color and light curve shape of the events.
Using the first version of the MLCS \citep{riess96} to model these relationships for SN Ia optical light curves, \citet{riess96b} analyzed the colors of 20 nearby SN Ia and found  $R_V = 2.6 \pm 0.3$, consistent with the Milky Way average.   \citet{tripp98} and \citet{trippbranch99} found $R_V \approx 1$, but they modeled intrinsic color and dust reddening as a single factor.  More recently, \citet{conley07}  found that the relation between SN Ia optical luminosity and apparent color, controlling for light curve shape, required a low value of $R_V \approx 1-1.7$, if the total color variation is interpreted as interstellar dust in the host galaxy.  An analysis of the color curves of SN Ia by \citet{nobili08} found $R_V \approx 1-1.7$.  \citet{hicken09b}  found that a dust absorption profile with $R_V = 1.7$ was favored using the MLCS2k2 model for the CfA3 sample \citep{hicken09a}.  These studies assumed that a universal color or dust absorption profile applied to all SN Ia.   Recently, \citet{wangx09b} separately fit $R_V \approx 1.6$ for a subset of SN Ia with high ejecta velocities, and $R_V \approx 2.4$ for a subset with normal ejecta velocities.  However, \citet{foleykasen11} find $R_V \approx 2.5$ for both subsets if the reddest SN are excluded.
For individual, highly extinguished SN with multi-wavelength coverage, $R_V$ can be fit precisely and values of 1.5-1.8 have been reported \citep{krisciunas07, elias-rosa06, elias-rosa07, wangx08}.  It has been suggested that low $R_V$ values could result from scattering by circumstellar dust in the local environment of the supernova \citep{lwang05, goobar08}.

 \citet{contreras10} recently presented the initial sample of nearby SN Ia light curves observed by the Carnegie Supernova Project (CSP), a subset of which were observed in the optical and near infrared.   \citet{folatelli10} compared the apparent colors of the SN to a subset suspected of having no contamination by dust, and found a dust law slope $R_V \approx 1.7$ best fit the SN sample, assuming a global value of $R_V$.  However, when the two reddest SN were removed from the sample, this global value changed to $R_V \approx 3.2$.  When minimizing dispersion in the Hubble diagram, they found $R_V \approx 1-2$, both with or without the reddest SN.   The different and unusual values of $R_V$ in the literature are problematic for the proper analysis and interpretation of SN Ia observables and for cosmological applications.

\citet{mandel09} presented a hierarchical Bayesian approach to constructing probabilistic models for SN Ia light curves.   This strategy was applied to the modeling of the extant near-infrared (NIR) light curves of SN Ia from PAIRITEL \citep{wood-vasey08} and the literature in the $JHK_s$ passbands.  Using an MCMC algorithm (\textsc{BayeSN}) designed specifically for hierarchical SN Ia light curve models, they computed coherent probabilistic inferences for individual supernovae and the population, taking into account multiple sources of randomness and uncertainty.   It was found that the variances of the peak absolute magnitudes were small, particularly in the $H$-band: $\sigma(M_H) \approx 0.11 \pm 0.03$ mag.  Since observations in the NIR passbands are insensitive to dust extinction, the estimation of host galaxy dust extinction was omitted from the NIR-only analysis in that paper.

In this paper, we expand upon the hierarchical modeling approach for SN Ia first described by \citet{mandel09}, and apply it to statistical modeling of SN Ia light curves in both the optical and near infrared, including the effects of host galaxy dust.  We describe a general mathematical representation of SN Ia light curves in terms of multiple decline rates over phase in different passbands.  This Differential Decline Rates representation is employed within a hierarchical model incorporating multiple random effects: measurement error, peculiar velocities, dust extinction, and intrinsic variation.  In particular, the intrinsic correlation structure of the light curves over phase and over wavelengths spanning optical to near infrared is explicitly modeled and estimated.  We also model the joint distribution of dust extinction and $R_V$.  To estimate the parameters of individual SN  and the characteristics of  the host galaxy dust distribution and intrinsic SN Ia populations, we have implemented a new \textsc{BayeSN} MCMC algorithm.  This new algorithm incorporates  enhancements to improve efficiency and convergence in the global parameter space.

We apply the hierarchical model to optical ($BVRI$) and NIR ($JH$) data of nearby SN Ia light curves from  the PAIRITEL, CfA3,  Carnegie Supernova Project (CSP) samples, and the literature.  We present inferences about the host galaxy dust population, and the correlation structure of the intrinsic SN Ia light curve population.   To check the fit of the hierarchical model to the sample of data,  we compare posterior predictive replications from the model  to the observed parameter distributions of colors, magnitudes and light curve shapes.   We quantify the utility of including NIR observations of SN Ia for improving estimates of host galaxy dust properties and distance predictions.  Analyzing optical and NIR light curves within the same model, we demonstrate that distance moduli to SN Ia observed with optical and near-infared light curve data can be predicted more accurately and precisely (\emph{rms} $= 0.11$ mag) than with optical data alone (\emph{rms} $=0.15$ mag).

This paper is organized as follows:  In \S 2, we describe our hierarchical Bayesian approach to constructing statistical models for SN Ia light curves.   In \S 3, we outline a new version of the $\textsc{BayeSN}$ algorithm for computing probabilistic inferences with the hierarchical model using the supernova data.  In \S 4, the application of the hierarchical model to nearby SN Ia data in the optical and near infrared is described, and posterior inferences about the dust and SN Ia light curve populations are summarized in \S 5.  In \S 6, we describe checks on our model inferences.  In \S 7 we employ cross-validation to construct a Hubble diagram of predicted distances to SN Ia, and demonstrate the advantages of including the NIR data for making more precise inferences about dust extinction and luminosity distances.   We conclude in \S 8.

In appendix \S \ref{appendix:ddr} we describe a non-parametric representation for individual light curves used by our model.   In appendix \S \ref{appendix:kcorr}, we describe the calculation of $K$-correction and Milky Way extinction effects based on spectral templates.  Hyperprior distributions are stated in appendix \S \ref{appendix:hyperpriors}.  Mathematical details of the new \textsc{BayeSN} algorithm are given in appendix \S \ref{appendix:bayesn}.

\section{Statistical Models for SN Ia Light Curves}
 
Inferences in supernova cosmology are based on statistical models  built from empirical data.
The application of statistical models for SN Ia to constraining the cosmological parameters has focused on using information in the apparent, optical light curves of SN Ia to infer their peak luminosities and estimate the luminosity distances.   In particular, these models capture empirical light curve width and color correlations with luminosity that allow SN Ia to be used as ``standardizable  candles'' \citep{ phillips93, hamuy96, phillips99, riess96, riess98, goldhaber01, prieto06, jha07, guy05, guy07}. 
Recent work has explored the utility of using spectral ratios or characteristics of spectral lines correlated with the luminosity to predict distances (\citealp{bailey09}; \citealp*{ blondin11}).  These methods show promise, but need to be validated on larger spectroscopic samples to be competitive with light curve methods.

The observed SN Ia light curve data is the result of the combination of multiple random effects.   Different ``normal'' SN Ia can have intrinsically distinct absolute light curves, peak luminosities and colors.  Each event can be extinguished and reddened by a different, and random, amount of host galaxy dust along the line of sight, and this dust may have different extinction laws as a function of wavelength for each event.    Before the light curve is recorded by an astronomer, it is subject to redshift effects, absorption due to Milky Way dust, and measurement error.  The measured redshift of each SN host galaxy is different from the cosmological redshift by a random peculiar velocity.   Sensible statistical models for SN Ia light curves must account for these multiple random effects in the data.

In the absence of the other effects, the apparent colors (e.g. $B - V$) of SN Ia at any phase are the sum of random intrinsic colors (e.g. $(B-V)_\text{int} \equiv M_B - M_V$) and random amounts of reddening by host galaxy dust ($A_B - A_V \equiv E(B-V)_\text{dust}$).  Hence, the joint distribution of the apparent colors over different wavelength ranges and at different phases is the convolution of the intrinsic color distribution and the dust reddening distribution.  Similarly, when the SN distances are known,  the extinguished absolute magnitudes (e.g. $V(t) - \mu$) at different wavelengths and phases are the sum of random intrinsic absolute light curves (e.g. $M_V(t)$) and random amounts of dust extinction (e.g. $A_V$) over wavelength.  The joint distribution of extinguished absolute light curves is the convolution of the intrinsic absolute light curve distribution and the dust extinction distribution.  Since dust extinction only makes objects appear dimmer and redder, the convolution with the dust distribution distorts the intrinsic distribution into the apparent distribution in the following ways.  Clearly, the apparent distribution will have a dimmer and redder average light curve.   The distribution of extinguished absolute magnitudes will be wider than the intrinsic distribution at any phase or wavelength.    The dust will also induce or increase apparent positive correlations between absolute magnitude and color (in the redder-dimmer sense), between two colors (redder-redder), and between absolute magnitudes (dimmer-dimmer) at different wavelengths and phases.  If we want to use SN Ia light curves to understand the statistical intrinsic properties of these physical events, or those of dust in distant galaxies, it is necessary to de-convolve these two effects in the observed data.  

Selecting a subsample of the observed SN, for example, the apparent blue end of the full sample, as representative of the ``intrinsic'' distribution, does not necessarily alleviate these distortions.   Some previous studies have selected an ``unreddened'' subsample based on auxiliary data, such as elliptical host galaxies or large physical separation of the SN from the center of the galaxy, which may suggest lack of dust extinction.   However, unless these auxiliary criteria can guarantee negligible dust effects, the resulting subsamples may still be distorted if there is a chance for some dust extinction.  \citet{hicken09b} showed with the large CfA3 sample \citep{hicken09a} that SN with moderate estimated dust extinction ($A_V \approx 0.4$ mag) are found in elliptical host galaxies or at large projected galactocentric distances between the host galaxy and the SN.  Furthermore, selecting an ``intrinsic'' subsample based on auxiliary data might distort inferences if the auxiliary properties are correlated with intrinsic properties, the distribution of which one is trying to identify.

Statistical errors in the estimates of the random effects can distort inferences of the intrinsic distribution.  For example, if distances to nearby SN Ia are estimated via recession velocities and the Hubble law to infer absolute magnitudes, the effects of random peculiar velocities on distance errors can distort the inferred intrinsic distributions of SN Ia absolute light curves.   Even in the absence of dust, a histogram of simple point estimates of peak absolute magnitudes for each SN (e.g. $V_0 - \mu(z)$) will appear broader than the true, intrinsic distribution, $P(M_V)$.  Similarly, random peculiar velocities can apparently induce or strengthen a positive correlation between the absolute magnitudes at two wavelengths, and distort correlations of absolute magnitudes with other observables if these random effects are not properly modeled.  Measurement errors and errors in dust extinction corrections will also tend to distort joint distributions of inferred variables.   Since extinction in the NIR is greatly diminished relative to the optical, inferences on the distribution of intrinsic SN Ia light curve properties in the NIR are much less vulnerable to distortions by dust, but are still affected by the other sources of error.  Supernovae far enough into the Hubble flow so that peculiar velocity effects are negligible will still be vulnerable to dust effects, especially at optical wavelengths.

Hierarchical Bayes provides a framework for the probabilistic modeling of multiple sources of randomness and uncertainty.  Its application to statistical modeling of SN Ia was first presented by \citet{mandel09}, who constructed hierarchical models for SN Ia light curves in the near infrared. The hierarchical framework provides a unified method of inference for populations and individuals of those populations.  It includes a population distribution that models intrinsic variations and correlations of SN Ia light curves, a population distribution for the the host galaxy dust to each SN, and models individual light curves, dust extinction, distances, and redshifts.  Using Bayes' Theorem, probabilistic estimates for the unknown parameters of individual SN, as well as the hyperparameters of the populations, can be computed coherently and consistently.

Statistical inference with hierarchical models provides a principled method of probabilistic deconvolution of physically distinct and random effects that are combined in the observed data.  Probabilistic inference allows for not only the estimation of each separate effect, but also the exploration of the joint uncertainties and trade-offs between the multiple effects.   It enables the estimation of the statistical characteristics of an underlying intrinsic population distribution while accounting for the distortions in the observed distribution caused by measurement error or other random effects.  Similar issues regarding  inferring the intrinsic distributions of inferred quantities in the presence of random error have been discussed and specific Bayesian techniques have been applied by \citet{bkelly07,bkelly07b}, \citet*{hogg10}, \citet{loredohendry10}, among others, in other astrophysical contexts.

Statistical modeling of SN Ia light curves is inherently a multi-dimensional problem.   Light curve observations are essentially noisy, usually irregular time series in multiple filters at different wavelengths.    However, the absolute light curves exhibit regularities: for example, the fast declining light curves tend to be intrinsically dimmer.  Existing models for optical light curves, e.g. MLCS  \citep{riess96, riess98, jha07}, SALT \citep{guy05, guy07}, and $\dmB$ templates \citep{hamuy96,prieto06} attempt to capture regularities by assuming strong functional forms, governing the ``global'' behavior of light curves over a wide range in phase and wavelength, and controlled by one or two parameters (e.g. $\Delta$; $x_1, c$; $\Delta m_{15}(B)$, respectively).
Although these formulations can be useful as a form of dimensionality reduction to project gross variations in high-dimensional data onto a small-dimensional latent parameter space, it is not clear that this reduction can be done cleanly without loss of statistical information contained in the light curves.  
Detailed studies of well-sampled light curves reveal that, for example, SN with the same $\Delta m_{15}(B)$ measurement (the magnitude change in the $B$-band light curve after 15 days from the peak) can display significant differences in their multi-band light curves over a range in phase \citep{folatelli10, hoflich10}, signifying that a single light curve shape parameter does not capture the full variety of light curve signals that are generated by the underlying explosion physics.  Furthermore, these global parameters lack direct interpretability:  even with a continuously well-sampled SN light curve, it is impossible to estimate the $\Delta$ parameter without knowing the particular templates that attempt to project it onto the latent parameter space.

In this paper, we take a different approach to modeling the light curves.  Instead of adopting a strong parameterization of the global behavior of the absolute light curves, we take a non-parametric approach that models the shape of the light curves ``locally.''  This does not mean that there are no parameters;  rather,  we use local parameters, describing the variations in signal in each neighborhood of phase and wavelength, to build up a model for the light curve over the full range of phase and wavelength.   We develop a Differential Decline Rates model (\S \ref{sec:rep}, \S \ref{appendix:ddr}) to represent the light curves in filters at multiple wavelengths using the decline rates over intervals in phase in each passband.   Regularities  underlying the population of light curves are then captured, not by one or two global parameters, but by inferring the correlations between the local parameters in the training set of well-sampled SN.    From this perspective, the light curves are modeled as stochastic processes with covariance structure over phase and wavelength that must be estimated.
The correlation structure in the light curves is modeled in the intrinsic population distribution for SN Ia light curves.  Although there may be many local parameters, they are not each statistically independent once the correlation structure of the population is learned.  Indeed, the intrinsic dimensionality of the light curves (i.e. the effective number of ``global'' degrees of freedom) is implicit in the estimated covariance structure, and does not need to be fixed \emph{a priori}.  
By incorporating this non-parametric light curve model in the hierarchical Bayes framework, we can coherently estimate the joint \emph{uncertainties} in the correlation structure and incorporate them into distance predictions for SN Ia.

The probabilistic hierarchical approach also provides a principled framework for dealing with missing data.  The observations are typically not obtained at an exactly regular cadence: observation times often can be random or clustered, with gaps in temporal coverage due to weather or instrumentation.   The SN in a given sample may not all be observed in the full set of passbands; in this paper, the SN are observed in the optical filters, but only a subset are observed in the NIR.  
The Bayesian approach deals with this by marginalizing over the unobserved light curves in the posterior distribution, thus incorporating this lack of information into inferences without omitting good incomplete data, which would be necessary if the analysis required the entire data set to be complete.   In the absence of complete data, the model makes the best estimates and predictions given the available observed data.

\begin{figure}[h]
\centering
\includegraphics[angle=0,scale=0.3]{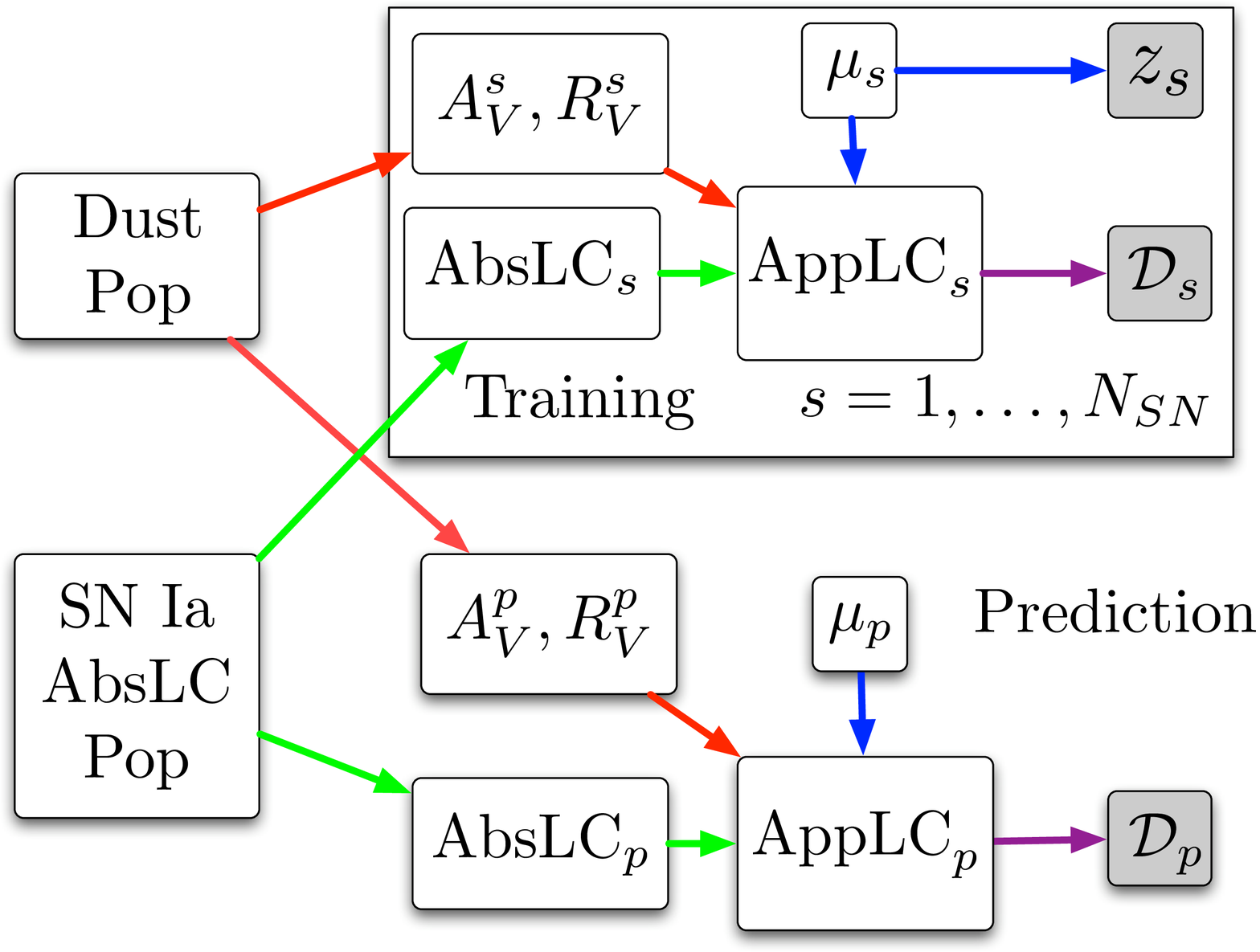}
\caption{\label{fig:dag}Hierarchical framework for statistical inference with SN Ia light curves.  The global posterior density of the hierarchical model parameters given the full SN data set is represented formally with a directed acyclic graph. Unknown parameters are represented by open nodes.  Observed data (redshifts $z$ and measured light curves $\mathcal{D}$) are represented by shaded nodes.  Each arrow or link describes a relationship of conditional probability.
The hierarchical model coherently incorporates randomness and uncertainties due to measurement error (purple), intrinsic SN variations (green), dust extinction and reddening (red), peculiar velocities and distances (blue) into inferences about individual SN and the population. The graph can be understood as a generative model for the data.  ``SN Ia AbsLC Pop'' represents parameters describing the population of SN Ia light curves, including intrinsic variations and correlations in shape, color and luminosity across multiple wavelengths.  From this population, each SN randomly draws a set of multi-wavelength light curves ``AbsLC''.  The box ``Dust Pop'' represents parameters governing the population distribution of host galaxy dust values.  Each SN randomly draws dust parameters $A_V, R_V$ from this distribution.  These dust parameters combine with the individual absolute light curves and distance modulus to generate an apparent light curve ``AppLC,''  which is sampled with noise to produce the observed multi-wavelength light curve data $\mathcal{D}$.  In the nearby universe, the distance modulus is related to the observed recession velocity or redshift through the Hubble law plus a noise term representing random peculiar velocities of host galaxies.  This random generative process is conceptually repeated for each SN in the data set.  The difference between ``training'' and distance prediction is that the latter does not condition on the redshift-distance likelihood information of the SN (bottom).}
\end{figure}

We have built upon the basic framework described by \citet{mandel09}.  The overall structure of the hierarchical Bayesian model is depicted by Fig. \ref{fig:dag}.   We describe each component of the model in turn.  

\subsection{Representation of Apparent Light Curves}\label{sec:rep}

An apparent light curve model at phase $t$ in rest-frame filter $F$ with parameters $ F_0$ and $\bm{\theta}^F = (  \bm{\theta}^F_\text{L}, \bm{\theta}^F_\text{NL} )$ is generally described by
\begin{equation}\label{eqn:genrep}
\begin{split}
\text{LC}^F(t ; F_0, \bm{\theta}^F) & = F_0 + l^F(t, \bm{\theta}^F) \\
&= F_0 + l^F_0(t; \bm{\theta}^F_\text{NL}) + \bm{l}^F_1(t ; \bm{\theta}^F_\text{NL}) \cdot \bm{\theta}^F_\text{L} .\\
\end{split}
\end{equation}
where $F_0$ is the apparent magnitude at $t = 0$ in rest-frame filter $F$, and $l^F(t; \bm{\theta}^F)$ is the normalized light curve in filter $F$, so that $l^F(0) = 0$.     The vector of linear light curve shape parameters is $\bm{\theta}^F_\text{L}$, and $\bm{\theta}^F_\text{NL} $ is a vector of non-linear light curve shape parameters.      The phase is defined in the rest-frame of the SN, with $t=0$ corresponding to the time of maximum light in $B$, $T_0$: $t = (T-T_0)/(1+z)$, where $z$ is the measured redshift, and $T$ is the time of observation.
For a multi-wavelength model using light curve observations corresponding to rest-frame filters $F^1, \ldots, F^N$, the apparent light curves of a SN are described by a vector of apparent light curve parameters 
\begin{equation}\label{eqn:phi}
\bm{\phi} = ( F^1_0, \ldots, F^N_0, \bm{\theta}^{F_1}, \ldots, \bm{\theta}^{F_N} )
\end{equation}
and the time of maximum light in $B$, $T_0$.

The models employed in this paper do not use non-linear shape parameters, so $\bm{\theta}^F = \bm{\theta}^F_\text{L}$.   To specify the light curve functions $l_0^F(t)$ and $\bm{l}_1^F(t)$, we construct a representation in terms of differential decline rates, as described in Appendix \S \ref{appendix:ddr}.   In this representation, the light curve shape parameters $\bm{\theta}^F$ for each filter are simply the changes in magnitude over disjoint intervals in phase.   Let $\bm{d}^F$ be a vector of decline rates of a light curve in filter $F$ on a grid in phase, set to $\bm{\tau} = (  -12,
   -8,
   -4,
   -2,
         0,
    2,
    4,
    6,
    8,
   10,
   12,
   15,
   18,
   23,
   30,
   37.5,
   45)$ days.  The decline rates are positive after maximum light and negative before peak in each filter.
For a given set of decline rates $\bm{\theta}^F = \bm{d}^F$, the normalized light curve for each filter at arbitrary phase is constructed with with a smooth curve defined by non-parametric regression cubic spline.  In this representation the light curve function $l_0^F(t) = 0$, and $\bm{l}_1(t)$ is determined by a linear smoothing spline: $\text{LC}^F(t ; F_0, \bm{\theta}^F) = F_0 + \bm{l}_1^F(t) \cdot \bm{\theta}^F$.  

\subsection{Likelihood Function for Apparent Light Curves}

In this section, we describe the likelihood function for the apparent light curve model parameters, conditional on the the observed light curve data.  The likelihood function explicitly accounts for $K$-corrections from the rest-frame filter to the observer-frame filter, Milky Way extinction, and photometric measurement error.

A light curve measurement in the observer frame $O$ at time $T$ is $m^O$.  This observation differs from the apparent light curve model in the rest-frame through $K$-corrections (to account for the redshifting of the SN spectrum),  Milky Way extinction, and measurement error.   At each redshift, we construct a unique mapping between each observer-frame filter $O$ and a rest-frame filter $F$ (in this paper, $O, F \in \{ B, V, R, I, J, H\}$).
\begin{equation}\label{eqn:single_obs}
\begin{split}
m^O &=  \text{kc}^{OF}(t; z, \bm{\phi}) + \text{gx}^{OF}(t; z, \bm{\phi}, E_{\text{MW}}) \\
&+ \text{LC}^F(t; F_0, \bm{\theta}^F) +\epsilon.
\end{split}
\end{equation}
The $K$-correction for the redshift $z$ supernova magnitude at phase $t$ from rest-frame filter $F$ to observer frame filter $O$ is  $\text{kc}^{OF}(t; z, \bm{\phi})$.  The $K$-correction has a dependence on the supernova spectral energy distribution (SED), and this is modeled as a function of apparent color.  Thus, it depends on the  apparent light curve parameters $\bm{\phi}$ only through the apparent model colors at the same phase.  For example, at low redshifts $z < 0.05$, if the observer-frame filter is $O = B$, then the rest-frame filter is $F = B$, and the dependence of $ \text{kc}^{OF}(t; z, \bm{\phi}) $ on the model light curve parameters $\bm{\phi}$ is through the apparent color $B-V$ at phase $t$.  The effective Milky Way extinction, $\text{gx}^{OF}(t; z, \bm{\phi}, E_\text{MW})$, also depends on the SN SED through the colors, and is also a function of the estimated color excess due to Milky Way dust,  $E_{\text{MW}} \equiv E(B-V)_\text{MW}$, which is obtained from the \citet*{sfd98} dust maps.   Details regarding the calculation of $K$-corrections and Milky Way extinction for SN Ia are presented in Appendix \S \ref{appendix:kcorr}.
The variance of the random error term $\epsilon$ includes photometric error, and estimated uncertainties in $K$-corrections and Milky Way extinction.

For each measurement in observer frame filter $O$ at observed time $T$, we can write down Eq. \ref{eqn:single_obs} relating the measurement to the rest-frame light curve model.    Let the observations in filter $O$ be arranged into a \emph{time-ordered} vector $\bm{m}^O$, with each observation listed from earliest to latest.   The corresponding equations can be also be time-ordered.  If the supernova light curve is observed in multiple filters, $O_1, \ldots, O_N$, we can arrange the full data in \emph{time-filter ordering}, so that $\bm{m} = ( \bm{m}^{O_1}, \ldots \bm{m}^{O_N})$, with the observer frame filters ranked from shortest to longest central wavelength.  With this arrangement, a time-filter ordered vector equation can be written for each SN:
\begin{equation}
\bm{m} =  \text{\bf KC}(T_0; z, \bm{\phi})+ \text{\bf GX}(T_0; z, \bm{\phi}, E_{\text{MW}}) + \bm{L}_2(T_0, z) \bm{\phi} + \bm{\epsilon}.
\end{equation}
We suppress the explicit dependence on the known observation times $T$.  Each of the terms depends on the time of maximum $T_0$, and the time-dilating redshift through the phase $t = (T-T_0)/(1+z)$.   Each element of the vectors $\text{\bf KC}$ and $\text{\bf GX}$ corresponds to the $K$-correction or Milky Way extinction scalar in Eq. \ref{eqn:single_obs}.
Here $\bm{L}_2$ is the unique matrix that, when multiplied with the apparent light curve parameter vector $\bm{\phi}$, computes the rest-frame apparent light curve model corresponding to each time-filter ordered observation in $\bm{m}$.  Its rows are constructed from the individual vectors $\bm{l}_1(t)$.

If the random errors $\bm{\epsilon}$ are normally distributed, the likelihood function of the unknowns $T_0$, $\bm{\phi}$  for the full data set for a single SN is
\begin{equation}\label{eqn:lc_lkhd}
\begin{split}
P( \bm{m} |& \, T_0, \bm{\phi}, z) = \\
N[\bm{m} &| \, \text{\bf KC}(T_0; z, \bm{\phi})+ \text{\bf GX}(T_0; z, \bm{\phi}, E_{\text{MW}}) \\
&+ \bm{L}_2(T_0, z) \bm{\phi}, \bm{W} ]
\end{split}
\end{equation}
where $\bm{W}$ is the error covariance matrix, and $N(\bm{x} | \bm{\mu}, \bm{\Sigma})$ is a multivariate Gaussian probability density with mean vector $\bm{\mu}$ and covariance matrix $\bm{\Sigma}$.  In general, $\bm{W}$ can be a full positive definite symmetric matrix, if the errors due to photometric calibration, $K$-corrections, and Milky Way extinction are correlated across the observer filters and observation times.   Our algorithms allow for this to be a full matrix, but for this paper, we take the simple approach of assuming it is diagonal, using only the measurement error variances.   We symbolize the light curve data, the observed magnitudes and times, and the error covariance for each SN, as $\mathcal{D}_s = \{ \bm{m}, \bm{W}, \bm{T} \}$.

\subsection{Redshift-Distance Likelihood Function}\label{sec:zmu}

The theoretical relation between the cosmological redshift $z_c$ and the luminosity distance $d_L$ to a SN in a smooth cosmological model depends on the cosmological parameters $\Omega_M, \Omega_\Lambda, w$ and the Hubble constant $h = H_o / 100 \text{ km s}^{-1}$.  At low redshifts, distances are insensitive to the cosmological model, and if we are concerned only with ratios of distances (or differences of distance moduli), then it is sufficient to fix $h$.  For this paper, we are not concerned with constraining cosmological parameters, but on the statistical properties of SN Ia light curves, so we fix $\Omega_M = 0.73, \Omega_\Lambda = 0.27, w = -1$ and $h = 0.72$.   
The cosmological redshift may differ from the measured redshift $z$ through measurement error and random peculiar velocities.  The expected value of the distance modulus at redshift $z$ is $f(z) = 25 + 5 \log_{10}[d_\text{L}(z) \, \text{Mpc}^{-1}]$.
As described in \citet{mandel09}, the likelihood function of the distance modulus given the measured redshift
\begin{equation}\label{eqn:z_lkhd}
P(\mu  |\, z) = N[\mu \, | \,  f(z) ,  \sigma^2_\mu = [f'(z)]^2(\sigma^2_z + \sigma^2_{\text{pec}}/c^2  ) ].
\end{equation}
In the low-$z$ regime, where $d_\text{L}(z)$ is linear in $z$ (the Hubble law), the variance is 
\begin{equation}\label{eqn:sigma_mu}
\sigma^2_{\mu} =  \left(\frac{5}{z \ln 10} \right)^2 \left[ \sigma_z^2 + \frac{\sigma_{\text{pec}}^2}{c^2} \right].
\end{equation}
where $\sigma_z^2$ is the redshift measurement variance and $\sigma_\text{pec}^2$ is the expected variance due to random peculiar velocities.  In this paper, we have alternately taken $\sigma_\text{pec} = 150$ \citep{radburn-smith04} and $300 \text{ km s}^{-1}$.  Our results are consistent between the two values.  We used the measured redshifts corrected to the cosmic microwave background frame and the local infall
flow model of \citet{mould00}.

In this paper, we are only concerned with evaluating distance predictions for low-$z$ SN Ia, so these parameters are fixed to their concordance values.
The dependence of the redshift-distance relation on the cosmological parameters could be made explicit by writing $P( \mu |\, z; \Omega_M, \Omega_\Lambda, w)$ and allowing them to be free parameters that appear in the global posterior density (Eq. \ref{eqn:globalposterior}) of a cosmological sample of SN Ia. 

\subsection{Latent Variable Model and Host Galaxy Dust}\label{sec:latent}

The vector $\bm{\phi}$ in Eq. \ref{eqn:phi} encodes the information needed to construct the apparent light curve model in the rest-frame filters.   Using the differential decline rates representation, this vector encodes the peak apparent magnitudes and the decline rates of the apparent light curve in multiple filters over intervals in phase.  The latent parameter vector $\bm{\psi}$ encodes the information for constructing the absolute light curve model in the rest-frame filters: the peak absolute magnitudes in rest frame filters, and the decline rates of the absolute light curves.  The two sets of parameters are related by host galaxy dust extinction and distance:
\begin{equation}\label{eqn:latent}
\bm{\phi} = \bm{\psi} +\bm{A}+ \bm{v} \mu.
\end{equation}
The vector $\bm{A} = A_V (\bm{\alpha} + \bm{\beta}/R_V)$ represents the effect of extinction on the absolute light curve parameters, and is a function of the host galaxy extinction, $A_V$, and the slope of the extinction law, $R_V$, using the dust extinction law of \citet*{ccm89}.   We model the effect of host galaxy extinction as described in \citet*{jha07}.  For a given $(A_V, R_V)$, the effective dust extinction in filter $F$ at phase $t$ is
\begin{equation}\label{eqn:AFt}
A_F(t) = A_V \zeta_F(t) (a_F + b_F/R_V).
\end{equation}
The coefficients $a_F$ and $b_F$ model the effect of dust extinction on the supernova SED within each passband $F$ at the time of maximum light.  The functions $\zeta_F(t)$ model the change of this effect with phase due to the evolving supernova SED.   The constant vector $\bm{\alpha}$ is constructed with components
\begin{equation}
\alpha_j = \begin{cases}  a_F, & \text{ if $\phi_j$ is a peak magnitude, $F_0$} \\
\Delta \zeta_F^k a_F, & \text{ if $\phi_j$ is a decline rate in filter $F$} \\
 & \text{ \,\,  between phases $\tau_k$ and $\tau_{k+1}$} 
\end{cases}
\end{equation}
where $\Delta \zeta_F^k \equiv [\zeta_F(\tau_{k+1}) - \zeta_F(\tau_k)]$.
The constant vector $\bm{\beta}$ is defined analogously, in terms of $\zeta_F$ and $b_F$.  This accounts for the effect of dust extinction on the apparent magnitudes and light curve shape through the evolving SN Ia SED with phase.

Since distance only changes the magnitude, but not the shape of the light curve (after accounting for time dilation), the constant vector $\bm{v}$ is defined with components
\begin{equation}
v_j = \begin{cases} 1, & \text{if $\phi_j$ is a peak magnitude, $F_0$} \\
0, & \text{otherwise} \end{cases}
\end{equation}
With these constructions, we use Eq. \ref{eqn:latent} to relate the apparent light curve parameters $\bm{\phi}$ to the latent variables of extinction $A_V, R_V$, the distance modulus $\mu$, and absolute (intrinsic)  light curve parameters 
\begin{equation}\label{eqn:psi}
\bm{\psi} = ( M^{F_1}, \ldots, M^{F_N}, \bm{\tilde{\theta}}^{F_1}, \ldots, \bm{\tilde{\theta}}^{F_N} )
\end{equation}
where $\bm{\tilde{\theta}}^{F}$ contain the decline rates of the absolute model light curves in each rest-frame filter.  Since the model for the extinction in each passband and phase, Eq. \ref{eqn:AFt}, is linear in the inverse of $R_V$, we find it useful to define $r_V = R_V^{-1}$ to simplify the notation.

\subsection{Population Distribution Model for Intrinsic Absolute Light Curves}\label{sec:snpop}

Even normal SN Ia do not all have the same luminosities, intrinsic colors, or light curve shapes.  The heterogeneities of these properties in the population of SN Ia -- which cannot be explained by dust or distance -- are called intrinsic variations.   Estimation of the covariances in the population of SN Ia light curves is crucial to the utility of SN Ia as standardizable candles for distance estimation.  For example,  the well-known width-luminosity correlation of optical light curves \citep{phillips93, hamuy96, phillips99} allows us to estimate the distance modulus to a SN to $\sim 0.2$ mag.

To model and capture intrinsic correlations between the absolute magnitudes at multiple wavelengths and the shapes of their light curves, we need to specify a general correlation structure for the population distribution of $\bm{\psi}$.  Ideally, the population distribution $P(\bm{\psi})$ would be specified by reliable astrophysical theory.  However, current explosion models for SN Ia do not provide such detailed guidance regarding the expected distribution of absolute light curve properties.  Thus, we seek to model the population distribution of light curves generally, and infer the statistical properties of the intrinsic variations from the data.  We capture the intrinsic variations and correlations of SN Ia absolute light curves by modeling the distribution of the intrinsic parameters $\bm{\psi}$ as a multivariate Gaussian:
\begin{equation}\label{eqn:population}
\bm{\psi}_s \sim N( \bm{\mu}_\psi, \bm{\Sigma}_\psi)
\end{equation}
with mean vector  $\bm{\mu}_\psi$ and intrinsic covariance matrix  $\bm{\Sigma}_\psi$.  The intrinsic covariance matrix models
population correlations between the peak absolute magnitudes at different wavelengths, correlations between the light curve decline rates at different phases and wavelengths, and correlations between peak absolute magnitudes and light curve decline rates  at different phases and wavelengths.   By capturing the population correlations of the absolute light curves in multiple filters at different wavelengths, we implicitly also model the correlation structure of the intrinsic colors.  The intrinsic covariance matrix can be readily decomposed into a matrix of intrinsic correlations, $\bm{R}_\psi$ and a vector of intrinsic standard deviations $\bm{\sigma}_\psi$, one for each component $\psi^i$:  $\bm{\Sigma}_\psi = \text{diag}(\bm{\sigma}_\psi) \bm{R}_\psi \, \text{diag}(\bm{\sigma}_\psi)$.   Each element of $\bm{R}_\psi$ is a correlation coefficient between $-1$ and $1$.  A valid $\bm{\Sigma}_\psi$ must  be positive definite and symmetric.  Since $\bm{\psi}$ directly describes the multi-band absolute light curves, Eq. \ref{eqn:population} models them as a stochastic (Gaussian) process.

Suppose the intrinsic parameters vector $\bm{\psi} = (\bm{\psi}^u, \bm{\psi}^o)$ of a particular SN can be partitioned into parameters $\bm{\psi}^o$ tightly constrained by its observations, and parameters $\bm{\psi}^u$ that are not directly observed.  For example, $\bm{\psi}^o$ may be the decline rates describing the shapes of this supernova's light curves, and $\bm{\psi}^u$ may be the peak absolute magnitudes of this SN under distance prediction.
With this population model for intrinsic light curve variations, it is simple to estimate $\bm{\psi}^u$ given $\bm{\psi}^o$.  The full hyperparameters can be partitioned in the same way:
\begin{eqnarray}
\bm{\mu}_\psi = \begin{pmatrix} \bm{\mu}_\psi^u \\ \bm{\mu}_\psi^o \end{pmatrix} , &
\bm{\Sigma}_\psi = \begin{pmatrix} \bm{\Sigma}_\psi^{uu} & \bm{\Sigma}_\psi^{uo}  \\ 
\bm{\Sigma}_\psi^{ou} & \bm{\Sigma}_\psi^{oo}  \end{pmatrix}
\end{eqnarray}
Using standard theorems for the multivariate normal distribution, the expected value of $\bm{\psi}^u$, conditional on  $\bm{\psi}^o$ is $\mathbb{E}[\bm{\psi}^u | \, \bm{\psi}^o] = \bm{\mu}_\psi^u + \bm{\Sigma}_\psi^{uo} \, ( \bm{\Sigma}_\psi^{oo} )^{-1} ( \bm{\psi}^o-  \bm{\mu}_\psi^o )$, and its conditional covariance (uncertainty) is $\cov[\bm{\psi}^u | \, \bm{\psi}^o] = \bm{\Sigma}_\psi^{uu} - \bm{\Sigma}_\psi^{uo}  ( \bm{\Sigma}_\psi^{oo} )^{-1} \bm{\Sigma}_\psi^{ou}$.  This example demonstrates how this model uses the correlation structure of the absolute light curves to relate inferred variables to observable quantities, and vice versa.  

In the absence of host galaxy dust ($A_V = 0$) and measurement error, an estimator for the predicted distance can be derived straightforwardly for well-sampled light curves.  Suppose that apparent light curve parameters $\bm{\phi}$ could be measured perfectly for well-sampled light curves with vanishing measurement error.  If the intrinsic mean $\bm{\mu}_{\psi}$ and covariance $\bm{\Sigma}_\psi$ were known, then the posterior prediction of the distance modulus has mean $\hat{\mu} = \hat{V}_\mu  \bm{v}^T \bm{\Sigma}_\psi^{-1} (\bm{\phi} - \bm{\mu}_\psi)$ and variance $\hat{V}_\mu = ( \bm{v}^T \bm{\Sigma}_\psi^{-1} \bm{v} )^{-1}$.  In fact, $\hat{\mu}$ is the minimum variance unbiased linear estimator of the distance modulus, a result that does not depend on the Gaussianity of the intrinsic distribution of $\bm{\psi}$, the absolute light curves parameters.    This can be shown by noting that $\hat{\mu}$ is the generalized least squares solution of Eq. \ref{eqn:latent} and invoking the Gauss-Markov theorem.
However, the presence of a random amount of host galaxy dust for each supernova, and finite sampling and measurement error of the light curves necessitates modeling these other aspects of the hierarchical structure.

Modeling intrinsic variations of $\bm{\psi}$ in the population using the covariance structure of a multivariate Gaussian is the simplest choice.  However, if non-Gaussianities become important then it will be possible to replace this simple assumption with more complex distributions.  For example, a Gaussian mixture model could be used to describe a multi-modal population, Student-$t$ distributions can be employed to model fat-tailed populations, and non-linear regression could capture non-linear correlation structure.   Alternatively, one might seek a representation or parameterization (\S \ref{sec:rep}) for $\bm{\psi}$ that makes its population distribution more amenable to modeling with a simple forms.   For the application in this paper, we did not find these more advanced approaches to be necessary, so we postpone their discussion for future work.

Observables that are not derived from light curve data, such as host galaxy masses \citep[e.g.][]{pkelly10,sullivan10} or spectroscopic measurements (e.g. \citealp{bailey09}, \citealp*{blondin11}, \citealp{foleykasen11}), can be correlated with the intrinsic absolute light curves.  They can be included in this framework by augmenting $\bm{\psi}$ with an auxiliary parameter, and specifying a likelihood function describing the uncertainty in the new observable.  The joint distribution $P(\bm{\psi})$ would model the covariance structure of the intrinsic light curves along with the auxiliary property, which can be leveraged to compute distance predictions using the extra information.

\subsection{Population Models for Host Galaxy Dust}\label{sec:dustpop}

We also adopt models for the population distribution of host galaxy dust parameters $A_V$ and $r_V$ for each SN.  Their joint population distribution can be factored as $P(A_V, r_V ) = P(r_V | \, A_V) P(A_V)$.    The extinction $A_V$ values are assumed to be drawn from an exponential distribution describing dust along lines of sight from SN host galaxies \citep*{jha07}:  $A_V \sim \text{Expon}(\tau_A)$.  The probability density is
\begin{equation}
P(A_V | \, \tau_A) = \begin{cases} \tau_A^{-1} e^{-A_V / \tau_A},  & A_V > 0 \\
0 & A_V \le 0 \end{cases}
\end{equation}
with an unknown hyperparameter, the exponential scale length $\tau_A$, which is inferred from the hierarchical posterior probability density conditional on the data.

Even along lines of sight within the Milky Way, interstellar dust can cause varying amounts of reddening for a given amount of absorption or extinction.  This ratio is captured by the parameter $R_V \equiv A_V / E(B-V)$.  Although the average value within the Milky Way is 3.1, it can range from $2.1$ to $5.8$, and depends on the nature of the dust grains \citep[][and references therein]{ccm89,draine03}.

Previous studies have focused on the treating $R_V$ as a constant for all supernovae, either set to the Milky Way value, or left as a fit parameter.  However, this is a rather strong assumption, so here we consider the possibility that the $R_V$ of dust within the distant host galaxies of SN Ia may vary within a common distribution.    We also wished to explore whether $R_V$ could be systematically different for SN with different $A_V$ dust extinctions.
To test for and capture potential population correlations between $A_V$ and $r_V$, we consider several models for the conditional population distribution $r_V \equiv R_V^{-1}$ given $A_V$.   We consider six cases with the following assumptions:

\begin{enumerate}

\item \bm{$R_V = 3.1$}.  The host galaxy dust law slope is fixed  to the Milky Way interstellar average for all SN.

\item {\bf CP}: (\bm{$R_V = \textbf{const}$}).  Complete pooling;  $R_V$ has the same value for each SN, but this value is unknown and inferred from the posterior density.

\item {\bf NP}.  No pooling. The $\{r_V^s \}$ for each SN are  completely independent with a flat prior $P( r_V^s ) \propto 1$ on each.

\item {\bf PP: m = 0}. Partial pooling.   The $\{r_V^s\}$ are conditionally independent draws from a common Gaussian population distribution independent from the magnitude of extinction $A_V$.   
\begin{equation}\label{rv_gauss}
r_V^s \sim N( \mu_r, \sigma^2_r)
\end{equation}
The mean $\mu_r$ and variance $\sigma^2_r$ of this population are unknown and inferred from the posterior density.

\item {\bf PP: m = 1}.  Partial pooling.   The $\{ r_V^s \}$ are conditionally independent draws from a common population distribution with a mean linearly dependent on  $A_V$.   
\begin{equation}
r_V^s | A_V \sim N( \beta_0 + \beta_1 A_V^s, \sigma^2_r)
\end{equation}
The regression coefficients $\bm{\beta}$ and residual variance $\sigma^2_r$ of this population are unknown and inferred from the posterior density.   The intercept $\beta_0$ represents the population mean $r_V$ value at vanishing extinction, and $\beta_1$ captures a potential trend of $r_V$ with increasing $A_V$.

\item {\bf PP: Steps}.  Partial pooling with the step function distribution.  The range in $A_V$ is divided up into two or four intervals (c.f. Tables \ref{table:Rv_2steps} \& \ref{table:Rv_4steps}).   Within each interval, the $r_V^s$ for each SN is a conditionally independent draw from a common Gaussian distribution with mean $\mu_r$ and variance $\sigma_r^2$ (Eq. \ref{rv_gauss}).  These hyperparameters are estimated from the posterior density.

\end{enumerate}
In all cases, we limited $R_V$ to an allowed range: $0.18 < r_V  < 0.7$ ($1.4 < R_V < 5.6$).   

Cases 4, 5 and 6 perform partial pooling, or shrinkage estimation, of the $r_V$ parameters, which is characteristic of hierarchical Bayes models.  In the \emph{complete pooling} case, it is assumed that the dust in the host galaxies of SN all have the same value of $R_V$, and thus all the information in the sample of SN is used (``pooled'') to infer the single $R_V$ value.  In the case of \emph{no pooling}, it is assumed that the host galaxy dust for different SN can have different, independent values of $r_V$, and that only the information from each SN is used to infer the $R_V$ value for that SN.  These two cases are limits of \emph{partial pooling}, in which the information for each SN is combined with that of the population to produce the individual $R_V$ estimates.  
The appropriate weight between the individual SN information and that of the population is negotiated by the inferred population variance, $\sigma^2_r$, and the precision with which $R_V$ can be estimated independently for each SN.  As $\sigma^2_r \rightarrow 0$, we obtain complete pooling, and for $\sigma^2_r \rightarrow \infty$, we have effectively no pooling.  At intermediate values, partial pooling finds a middle ground between the noisy and possibly unstable estimates of no pooling, and the possibly unrealistic strong assumptions of complete pooling.   The hyperparameter  $\sigma^2_r$ can be understood as the residual variance  of $r_V$ in the sample after accounting for the other sources of error for each individual SN.
Shrinkage accounts for the fact that a histogram or scatter plot of individual, simple, point estimates of derived quantities will be wider than the true, intrinsic distribution of those quantities, if those point estimates have significant uncertainties, and it reduces the mean squared error of each parameter estimate.  From a non-Bayesian perspective, shrinkage can be regarded as an adaptive regularization that determines from the data how much to allow an individual estimate to deviate from the population average or trend.   Shrinkage estimation with multi-level models has been discussed recently by \citet{loredohendry10}.

We can test the hypothesis that the dust law parameter $r_V$  has no dependence on host galaxy extinction $A_V$ by comparing the results from fitting Cases 4 and 5.  Since Case 4 is a nested case of Case 5 with $\beta_1 = 0$, we can check to see whether or not the marginal posterior density of $\beta_1$ is consistent with zero when fitting Case 5.  Similarly, with Case 6, we can check to see whether the population means $\mu_r$ in each interval in $A_V$ are consistent with each other across the range of $A_V$, or if there are significant differences.

For brevity and specificity, in subsequent sections describing statistical computation, we adopt the assumptions of Case 5; the hyperparameters governing the $r_V$ population are $\bm{\beta}$ and $\sigma^2_r$.  For other cases, the hyperparameters are $\bm{\mu}_r$ and $\sigma^2_r$ and the algorithms are appropriately modified to account for the different models.

\subsection{Specifying the Hyperpriors}

Diffuse, or non-informative, hyperpriors are adopted on the highest-level hyperparameters of the hierarchical model: $\bm{\mu}_\psi, \bm{\Sigma}_\psi$ for the SN light curve population and $\tau_A$, $\bm{\mu}_r$ or $\bm{\beta}$, and $\sigma_r^2$ for the host galaxy dust population.   As the number of well-observed supernovae, $\nsn$, becomes larger, the influence of the hyperpriors on the posterior estimates of the hyperparameters diminishes.   Thus, so long as we include a sufficient number of SN in the hierarchical model, we can choose the particular, analytic forms of diffuse  hyperpriors for computational convenience.
We discuss some mathematical details in the appendix (\S \ref{appendix:hyperpriors}).

\subsection{Global Posterior Probability Density}

\citet{mandel09} derived the global posterior probability density over all the SN in the training set as a product of the light curve and redshift-distance likelihood functions for individual SN, population distributions for SN Ia light curves and host galaxy dust, and the hyperpriors on the hyperparameters of the population distributions.  We construct the global posterior using the new component probability models described above.  Let the time-filter ordered light curve data in the observer frame for SN $s$ be $\mathcal{D}_s$, with measured redshift $z_s$. The unknown parameters for an individual SN are the apparent light curve parameters $\bm{\phi}_s$, the distance modulus $\mu_s$, the extinction $A_V^s$, the slope of the dust law $r_V^s$, and the time of maximum $T_0^s$.   For a given set of population hyperparameters, $\bm{\mu}_\psi, \bm{\Sigma}_\psi, \tau_A, \bm{\beta}, \sigma^2_r$, the conditional posterior density for an individual SN is
\begin{equation}\label{eqn:singlesn}
\begin{split}
P&( \bm{\phi}_s , T_0^s, \mu_s, A_V^s, r_V^s | \, \mathcal{D}_s, z_s;  \bm{\mu}_\psi, \bm{\Sigma}_\psi, \tau_A, \bm{\beta}, \sigma^2_r) \\ &\propto 
P( \bm{m}_s |\, T_0^s, \bm{\phi}_s, z_s) \times P(\mu_s | \, z_s) \\
&\times P(\bm{\psi}_s =  \bm{\phi}_s  - \bm{v} \mu_s - \bm{A}_s | \, \bm{\mu}_\psi, \bm{\Sigma}_\psi) \\ &\times P(A_V^s, r_V^s |\, \tau_A, \bm{\beta}, \sigma^2_r) .
\end{split}
\end{equation}

The training set data is comprised of the light curve data for all the SN in the training set $\mathcal{D} = \{ \mathcal{D}_s \}$ and their redshifts $\mathcal{Z} = \{ z_s \}$.   The unknown hyperparameters of the populations are the mean and covariance of the absolute light curve parameters $\bm{\mu}_\psi, \bm{\Sigma}_\psi$, the exponential scale of the extinction distribution $\tau_A$, and the hyperparameters describing the $R_V^{-1}$ distribution.
The global joint posterior density of all supernova observables $\{ \bm{\phi}_s \}$, distance moduli $\{\mu_s\}$, 
dust parameters $\{A_V^s, r_V^s\}$, and the population hyperparameters conditioned on the database $\mathcal{D}, \mathcal{Z}$  is proportional to 
the product of $N_{\text{SN}}$ individual densities multiplied by the hyperpriors:
\begin{equation}\label{eqn:globalposterior}
\begin{split}
P&(\{ \bm{\phi}_s, T_0^s, \mu_s, A_V^s, r_V^s\} ;  \bm{\mu}_\psi, \bm{\Sigma}_\psi, \tau_A, \bm{\beta}, \sigma^2_r| \, \mathcal{D}, \mathcal{Z}) \\
&\propto \Bigg[ \prod_{s=1}^{N_{\text{SN}}} P( \bm{m}_s |\, T_0^s, \bm{\phi}_s, z_s) \times P(\mu_s | \, z_s) \\
&\times P(\bm{\psi}_s =  \bm{\phi}_s  - \bm{v} \mu_s - \bm{A}_s | \, \bm{\mu}_\psi, \bm{\Sigma}_\psi) \\ &\times P(A_V^s, r_V^s |\, \tau_A, \bm{\beta}, \sigma^2_r)  \Bigg]  \times P(\bm{\mu}_\psi, \bm{\Sigma}_\psi) \times P(\tau_A, \bm{\beta}, \sigma^2_r).
\end{split}
\end{equation}
To predict the distance of a SN using its light curve data, one sets the redshift-distance likelihood $P(\tilde{\mu}_s | \, \tilde{z}_s) \propto 1$ in Eq. \ref{eqn:singlesn}, where we use tilde on parameters and data for prediction set SN.   The tilde redshift $\tilde{z}_s$ is used for time-dilation and $K$-correction in the light curve likelihood function, but not to constrain the distance modulus in the redshift-distance likelihood.  The marginal posterior predictive density is $P( \tilde{\mu}_s | \, \tilde{D}_s, \tilde{z}_s; \mathcal{D}, \mathcal{Z})$, obtained by integrating Eq. \ref{eqn:globalposterior}.

Equation \ref{eqn:globalposterior} is an explicit statement of the objective function for statistical inference for both training and prediction with the hierarchical model.   By computing Eq. \ref{eqn:globalposterior}, we solve for the most likely distributions of host galaxy dust and intrinsic luminosities, colors and light curve shapes that best account for the apparent distributions, conditional on the model assumptions.
A directed acyclic graph (DAG) representing the hierarchical model and the global posterior density is shown in Figure \ref{fig:dag}.  The graph depicts a generative probabilistic model linking together the populations and individuals, and parameters and hyperparameters to the data \citep{mandel09}.
For simplicity, we have not shown the dependence of the light curve likelihood function on redshift through time-dilation and $K$-correction, and only show the redshift-distance dependence, which is the key difference between training and prediction.

\section{Improved MCMC with \textsc{BayeSN}}\label{sec:bayesn}

It is important to distinguish between the tasks of statistical inference and statistical computation.  The former entails deriving estimators for unknown quantities, conditional on data and the assumptions of the statistical model.  We have done that in the previous section by deriving the global posterior probability density, Eq. \ref{eqn:globalposterior}.  The task of statistical computation is accomplished by specifying, constructing, and implementing algorithms for computing the numerical values of these estimators for the observed values of the data, under the assumptions of the model.  In this section, we describe our strategy for statistical computation of the posterior estimates by stochastic sampling from the global posterior probability density,
Eq. \ref{eqn:globalposterior}.

We perform fully Bayesian inference of the hierarchical model using Markov Chain Monte Carlo (MCMC) sampling methods such as the Metropolis-Hastings algorithm \citep{metropolis53,hastings70} and Gibbs sampling \citep{geman84}.  \citet{mandel09} presented the first MCMC algorithm (\textsc{BayeSN}) for hierarchical Bayesian inference with supernova light curves.  We have made many modifications to the original \textsc{BayeSN} algorithm to incorporate the modeling of host galaxy dust, $K$-corrections from the rest-frame model light curves to the observer-frame measured magnitudes, and to improve the computational speed and convergence rate of the chains.  The new algorithm is largely comprised of Gibbs sampling and Metropolis-Hastings jumps that do not require the user to ``tune'' jump sizes for each SN, as would be required for ordinary Metropolis steps.     Instead, the algorithm only requires the specification of a jump size for the scalar time of maximum for each SN, $T_0^s$, which is a relatively easy task: a \emph{rms} jump proposal of $\sim 0.5$ days was generically successful for rapid convergence for all SN.   Movement along the other dimensions of parameter space is accomplished by Gibbs sampling or more general Metropolis-Hastings proposals that exploit the conditional structure of the posterior distribution to adaptively propose more global moves.  By minimizing the amount of manual tuning required before running the MCMC, we have increased the ease-of-use and practical applicability of MCMC for fitting comprehensive hierarchical models for SN Ia.  A basic introduction to MCMC algorithms for SN Ia probabilistic inference was given by \citet{mandel09};  statistical references include \citet{liu02, gelman_bda, robert05}.

We use this new \textsc{BayeSN} code to sample from the global posterior probability distribution of all individual parameters and the population hyperparameters of SN Ia light curves and host galaxy dust, conditioned on the supernova light curve data and redshifts.
Here, we sketch the improved \textsc{BayeSN} Gibbs sampling algorithm.  Further mathematical details can be found in Appendix \S \ref{appendix:bayesn}.
The state of the chain is determined by the current values of all the parameters and hyperparameters:
\begin{equation}
\bm{\mathcal{S}} = ( \{\bm{\phi}_s, T_0^s, \mu_s,  A_V^s, r_V^s\}, \bm{\mu}_\psi,  \bm{\Sigma}_\psi, \tau_A, \bm{\beta}, \sigma^2_r ).
\end{equation}
We generate random samples from the global posterior distribution $P(\bm{\mathcal{S}} | \, \mathcal{D}, \mathcal{Z})$ using a Markov chain, conditioned on the photometric light curve data for all SN $\mathcal{D} = \{ \mathcal{D}_s \}$, and their redshifts $\mathcal{Z} = \{ z_s \}$.
We begin with crude, randomized guesses of the individual supernova parameters $\{ \bm{\phi}_s, T_0^s, \mu_s,  A_V^s, r_V^s \}$ for all supernovae $s$ in the data set.  In each step, we update a subset, or block, of parameters  from their conditional posterior density with the complement set of parameters (and the data) fixed to their current values.  The choice of parameter blocks exploits the conditional independence structure of the directed acyclic graph of the inference, Fig. \ref{fig:dag}.

\begin{enumerate}

\item  Compute the absolute light curve parameters $\{\bm{\psi}_s\}$ using Eq. \ref{eqn:latent}.  The conditional posterior density of the light curve hyperparameters is $P(\bm{\mu}_\psi, \bm{\Sigma}_\psi | \, \cdot, \mathcal{D}, \mathcal{Z} )  = P(\bm{\mu}_\psi | \,  \bm{\Sigma}_\psi;  \{\bm{\psi}_s \}) P( \bm{\Sigma}_\psi | \, \{\bm{\psi}_s \})$.  We update $\bm{\Sigma}_\psi$ from the second factor, and then update $\bm{\mu}_\psi$ given $\bm{\Sigma}_\psi$ from the first (see \S \ref{appendix:bayesn} for details). (We use the notation $(\cdot)$ to indicate all the other parameters and data that have not been explicitly indicated.)

\item Draw a new extinction scale hyperparameter $\tau_A$ from the conditional distribution $P( \tau_A | \, \cdot, \mathcal{D}, \mathcal{Z} ) = P( \tau_A | \, \{A_V^s\})$, by drawing a random number from an inverse gamma distribution.

\item Obtain new values of  $\bm{\beta}$ and $\sigma^2_r$ from $P(\bm{\beta}, \sigma^2_r | \, \cdot, \mathcal{D}, \mathcal{Z} ) = P(\bm{\beta} | \, \sigma^2_r, \{r_V^s, A_V^s\} ) P(\sigma^2_r | \, \{r_V^s, A_V^s\})$.  First draw a new $\sigma^2_r$ from an inverse gamma distribution.  Conditional on that value, draw a new $\bm{\beta}$ from a Gaussian.

\item Repeat the following steps for each supernova $s$.
\begin{enumerate}

 \item  Propose a new time of maximum $T_{0,s}^* \sim N(T_{0,s}, s_T^2)$ according to a random walk from the previous estimate.  
It is usually sufficient to use $s_T \approx 0.5$ days. 
 Given $T_{0,s}^*$, propose a new set of apparent light curve parameters for all filters, $\bm{\phi}^*$, from the distribution in \S \ref{appendix:bayesn}.  Compute the Metropolis-Hastings ratio $r$ to decide whether to accept the joint proposal $(T_{0}^*, \bm{\phi}^*)$ or to stay at the current values $(T_{0}, \bm{\phi})$.

\item Update the new distance modulus $\mu_s$ from the conditional probability density $P(\mu_s | \, \cdot, \mathcal{D}_s, z_s) = P( \mu_s | \, \phi_s, A_V^s, r_V^s; \bm{\mu}_\psi, \bm{\Sigma}_\psi; z_s)$, by sampling from a Gaussian distribution.

\item Draw a new host galaxy extinction $A_V^s$ from the conditional posterior probability density $P(A_V^s | \, \cdot, \mathcal{D}_s, z_s) = P(A_V^s | \bm{\phi}_s, \mu_s, r_V^s; \bm{\mu}_\psi, \bm{\Sigma}_\psi, \tau_A, \bm{\beta}, \sigma^2_r)$.  This can be shown to be a truncated Gaussian distribution in $A_V^s > 0$.  

\item Draw a new host galaxy dust law slope $r_V^s$ from the conditional posterior $P(r_V^s | \, \cdot, \mathcal{D}_s, z_s) = P(r_V^s |  \bm{\phi}_s, \mu_s, A_V^s; \bm{\mu}_\psi, \bm{\Sigma}_\psi, \bm{\beta}, \sigma^2_r)$ by evaluating the pdf on a fine grid on $0.18 < r_V  < 0.7$ and using griddy Gibbs sampling \citep{ritter92}.

\item (optional) Perform a translation in the space of distance versus extinction: $(A_V^s, \mu_s) \rightarrow (A_V^s, \mu_s) + \gamma (1, -x)$.  Here, $x$ determines a random direction along the trade-off.   We select a random $\gamma$ using generalized conditional sampling \citep{liu02}, and then translate to update $(A_V^s, \mu_s)$.

\end{enumerate}
\item Finally, record the current state of the chain $\bm{\mathcal{S}}$.  Repeat all steps until convergence.
\end{enumerate}
This algorithm generates an irreducible and ergodic Markov chain that will converge to a stationary distribution equal to the the global posterior density, Eq. \ref{eqn:globalposterior}, according to standard theorems \citep[e.g.][]{robert05}.  It converges without Step 4e, but this step speeds the exploration of the trade-off between extinction and distance for each SN.  We found that including this step reduces the auto-correlation scale for the slowest converging $A_V^s$  by a factor of $\sim 3$.  

\section{Application}

\subsection{Data Sets}

In this section, we describe the application of the hierarchical framework to inference with nearby SN with optical and NIR light curve observations ($BVRIJH$).

\citet{mandel09} analyzed a sample of SN with near infrared $JHK_s$ light curve observations compiled from the the PAIRITEL sample \citep{wood-vasey08}, and available published light curves from the literature  \citep{jha99,hernandez00, krisciunas00, krisciunas01, dipaola02, valentini03, krisciunas03, krisciunas04b, krisciunas04c, krisciunas07,elias-rosa06, pastorello07a, stanishev07, pignata08}.  Nearly all of the SN in the PAIRITEL sample were also observed in optical filters $UBVRI$ or $UBVr'i'$ by \citet{hicken09a}, and nearly all of the SN from the literature were also observed at optical wavelengths.    We selected the same set of SN with NIR light curves used in \citet{mandel09} with the following exceptions:  SN 2005ao was omitted for lack of quality data points, and  SN 2006lf was omitted because its position was near the Galactic plane, so its Milky Way dust reddening estimate was very large and unreliable.  Extensive studies of two of the well-sampled supernovae, SN 2005cf and SN 2006X, were presented by \citet{wangx08,wangx09}.

Since the number of SN Ia with NIR light curves is still small, we supplemented this set  with SN from the recent CfA3 sample of nearby SN Ia \citep{hicken09a} that had well-sampled optical light curves.   The additional light curves increase the statistical strength in the optical bands and stabilize estimation of the full hierarchical model.  They also provide a set of optical-only light curves to compare against the set of SN with optical and NIR light curves (\S \ref{sec:prediction}). We included SN with high quality light curve data at  $0.01 < z < 0.065$, with more than five observations in $B$ band, and with the first observation in $B$ occuring less than 10 days past maximum.  Since these light curves lack NIR data, the model marginalizes over the missing light curves for all inferences.  Some of the SN in the CfA3 sample were observed in the $RI$ passbands, while some were observed in the $r'i'$ passbands.  In both cases, we map these observer-frame passbands to rest-frame $RI$ passbands, and in the latter case, the $K$-correction takes into account the cross-filter transformation.  The $K$-corrections for observations in $B$ and $V$ filters mapped them to rest-frame $B$ and $V$ filters.  The $J$ and $H$ observations were mapped to rest-frame $J$ and $H$ filters defined by 2MASS.

We use only normal SN Ia with $B$ decline rates $0.75 < \Delta m_{15}(B) < 1.6$.  As discussed by \citet{hicken09a}, fast decliners and SN 1991bg-like objects have different luminosity-light curve shape relations than normal SN Ia, and should be modeled separately, so we do not include them in our analysis.  We have also excluded the peculiar SN 2006bt from the sample \citep{foley10_06bt}.  The full ``CfA+literature'' sample consists of 110 supernovae, 37 of which have NIR observations.

We modeled the rest-frame $BVRIJH$ light curves of this set of SN listed in Table \ref{table:sn_lcdust}.   For each SN, if there was data in an observer frame passband mapping to a given rest-frame filter, we list the fitted peak apparent magnitude in the rest-frame filter.  If there were no observations for a given rest-frame passband, then the estimated peak magnitude is not listed.  We have omitted ultraviolet data for the analysis in this paper.  
\citet{kessler09} found that differences in the $U$-band model between MLCS2k2 and SALT2 lead to large ($\Delta w  \sim 0.2$) differences in the cosmology using high redshift samples. As this work focuses on low redshift data, we have omitted the $U$ data to avoid calibration and standardization problems, selection effects, and $S$-corrections between the $u$ and $U$ bands. Future work, applying our framework to high redshift SN Ia samples for cosmological analysis will carefully incorporate $U$-band data.

\citet{contreras10} recently published light curve observations for a set of nearby SN as part of the Carnegie Supernova Project, and an analysis of this data was discussed by \citet{folatelli10}.  These are high-quality well-sampled light curves with optical coverage, and a subset included contemporaneous NIR observations as well.   Comparison of the CSP and PAIRITEL data reduction and calibration for SN observed in both samples is ongoing.   For  joint analysis of the current published photometry of these sets, in \S \ref{sec:prediction}, we augment our sample with 27 SN with CSP data passing the criteria and examine the distance predictions for each set as a consistency check.  Twenty of these SN have joint optical and NIR light curve observations.

\subsection{Statistical Computation}

We ran the new \textsc{BayeSN} code (\S \ref{sec:bayesn}) to coherently fit the apparent light curves with the Differential Decline Rates model (\S \ref{sec:rep}), to estimate host galaxy dust extinction (\S \ref{sec:latent}) and distance moduli (\S \ref{sec:zmu}), and to infer intrinsic variations and correlations (\S \ref{sec:snpop}) and the host galaxy dust population (\S \ref{sec:dustpop}).  The code samples the global posterior density (Eq. \ref{eqn:globalposterior}) over all unknowns given the data.

We seeded each chain with random, initial values of the SN and dust parameters  $\{\bm{\phi}_s, A_V^s, \mu_s\}$.
In many cases, the $B$ light curve was sufficiently well-sampled that the time of maximum could be unambiguously determined.  In these cases, we fixed the $T_0$ to that value and did not re-estimate it when running the sampler (by setting $s_T = 0$ in step 4a).    For SN with more uncertain $T_0$ we re-estimated it during the fitting ($s_T = 0.5$d).   Under Case 1 (\S \ref{sec:dustpop}), the initial values of $R_V$ were all fixed to 3.1.  For all other $r_V$ cases, the initial $\{ r_V^s \}$ were randomized.

\begin{figure}[t]
\centering
\includegraphics[angle=0,scale=0.45]{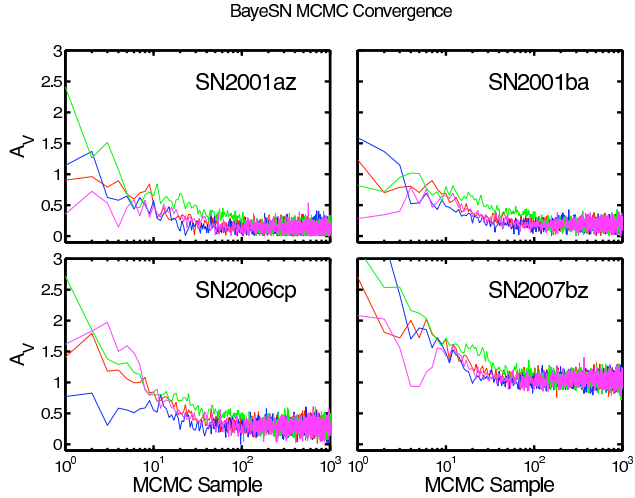}
\caption{\label{fig:Av_conv}Example sample paths of Markov Chain Monte Carlo (MCMC) chains generated by the \textsc{BayeSN} MCMC sampling code.  The full chain stochastically samples the parameter space of all individual SN in the set, and the populations of SN Ia light curves and the dust.  This plot focuses on the coordinates of the chain concerning the visual extinction $A_V$ to  particular SN.  Each color represents an independent chain starting from a randomized initial guess.  The chains explore the full parameter space and converge within a few hundred iterations upon the same global posterior distribution.  The posterior uncertainty in the estimate is reflected in the distribution (variability) of the chain samples upon convergence.  The plot depicts the simultaneous convergence of the chains, both for the estimate of a single SN, and for estimates of the ensemble of SN, ensuring the attainment of a consistent global solution for the SN population.  Each color represents one of four independent chains.  For example, the blue line in each panel is a different coordinate (projection) of the \emph{same} MCMC chain.}
\end{figure}

The initial dust configuration of the SN set is random, and we do not make an initial estimation of the $A_V^s$ value before running \textsc{BayeSN}.  Thus, it is possible that this randomized initialization of dust values may assign a high $A_V^s$ to an apparently blue SN, and a low $A_V^s$ to an apparently red object.  If we see that multiple chains starting with different and random initial dust configurations eventually converge to the same posterior estimate, we can be reassured that our final inferences are independent of the initial assignments of $A_V, r_V$ values, and that the probabilistic inference has sorted out the probable dust values over the set of SN.   In Figure \ref{fig:Av_conv}, we show that four independent and parallel chains training the hierarchical model over the set of SN, each starting with a different initial value for the host galaxy extinction, converge to the same final estimates in the long run of the MCMC.  This shows that our final inferences for the trained model are robust to the initial values of host galaxy dust, and indicates the self-consistency of estimates and convergence of the algorithm to a unique joint solution over the full set of SN.

To perform training and prediction, the \textsc{BayeSN} code generated four parallel, independent chains of $2 \times 10^4$ complete cycles (Steps 1-5).  
The initial values for each chain were generated by using different random numbers for each independent chain.
We thinned out the chains by recording only every 40th value.  This reduces the autocorrelation between successive recorded samples and saves memory.  To assess convergence, we computed the Gelman-Rubin statistic \citep{gelman92} for each parameter in the chain to compare the coverages of the independent chains.  We considered a maximum G-R ratio less than 1.10 to indicate convergence.
We discarded  the first 20\% of each chain as burn-in, and the chains were concatenated for analysis.

\section{Results: Posterior Inferences}

In this section, we report the posterior inferences of light curves and the population when the training set consists of all the SN and their redshifts $(\mathcal{D}, \mathcal{Z})$.   We report the posterior inference obtained when adopting Case 5 ($m=1$) for the $(A_V, r_V)$ population model, which models linear trends between the dust slope $r_V$ and the dust extinction $A_V$.  Posterior inferences can be described in terms of light curve fits and dust estimates for individual SN, intrinsic covariances in the population of SN light curves, and the population distribution and correlations of host galaxy dust properties.

\subsection{Individual Supernovae}\label{sec:individuals}

\begin{figure}[t]
\centering
\includegraphics[angle=0,scale=0.35]{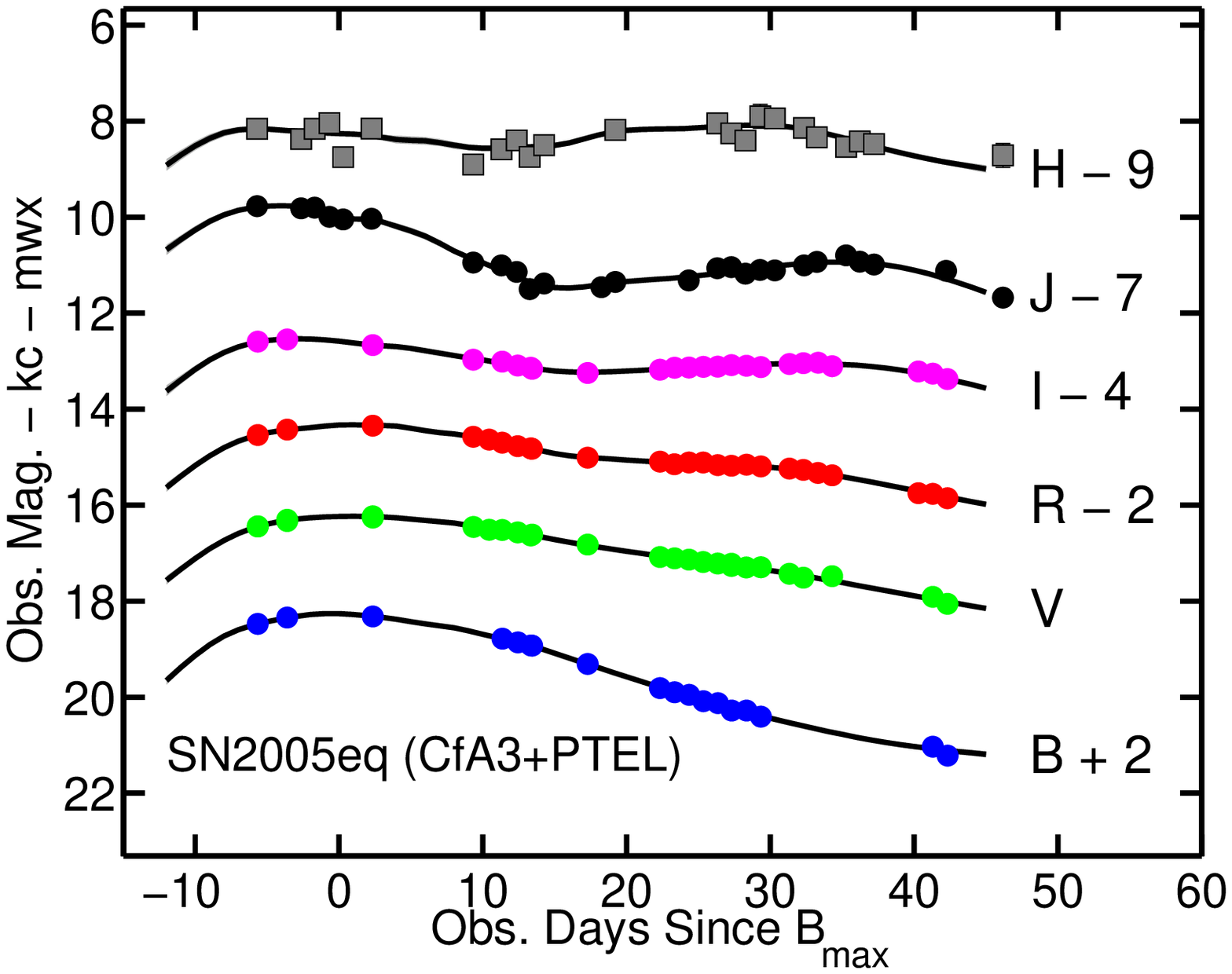}
\vfill
\includegraphics[angle=0,scale=0.35]{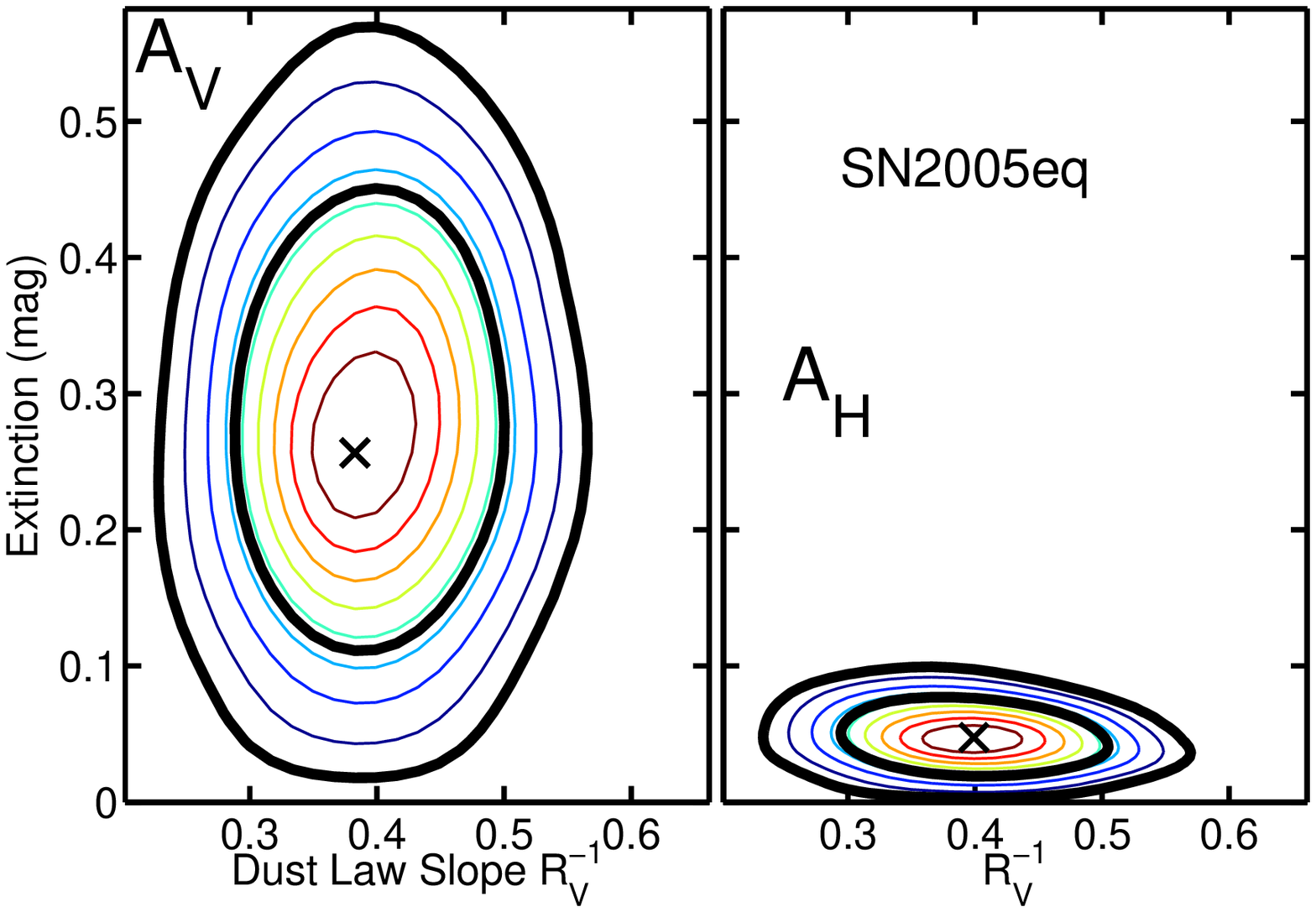}
\caption{\label{fig:sn05eq}(top) Optical \citep[CfA3,][]{hicken09a} and NIR \citep[PAIRITEL,][]{wood-vasey08} observations of nearby Type Ia SN 2005eq are fitted with a multi-band light curve model.    The points are the observed magnitudes in each filter minus estimated $K$-corrections and Milky Way extinction.  (bottom) Optical  and NIR light curves of SN 2005eq are used to infer the host galaxy dust extinction properties. The hierarchical model enables coherent inference of host galaxy dust properties $(A_V, R_V)$ (assuming a CCM dust law), while marginalizing over the posterior uncertainties in the dust and SN light curve populations.  The cross indicates the marginal bivariate mode, and the two black contours contain 68\% and 95\% of the posterior probability. The inferred NIR extinction $A_H$ is much smaller than the optical extinction $A_V$ and has much smaller uncertainty.  This SN exhibits moderate extinction and reddening due to host galaxy dust.}
\end{figure}

Optical and NIR light curve fits in the rest-frame are shown for one supernova, SN 2005eq, in  Fig. \ref{fig:sn05eq}.    The points are the measured magnitudes in the observer frame minus the estimated $K$-corrections and Milky Way extinction in each passband.  The black curves represent the fitted apparent light curves in each rest-frame passband, with each light curve represented by the differential decline rates model (\S \ref{appendix:ddr}).   The peak apparent magnitudes for each SN and the decline rate $\Delta m_{15}(B)$ are listed in Table \ref{table:sn_lcdust}.

We also depict the posterior inferences of the dust properties: the visual extinction $A_V$, the NIR extinction, $A_H$, and the slope of the extinction law $r_V \equiv R_V^{-1}$.  The bivariate marginal probability densities were estimated from the MCMC samples using kernel density estimation.  The marginal distributions integrate over the posterior uncertainties in individual light curve fits and the population distribution.
For SN 2005eq, we find a moderate amount of visual extinction, $A_V \sim 0.3$ mag.  We can see from the side-by-side comparison that not only is the $H$-band extinction about five times smaller, but its uncertainty is also much smaller.  

Since dust extinction is nonnegative, $A_V \ge 0$, the posterior probability densities of the dust parameters is highly non-gaussian for SN with low extinction.
For example, from Table \ref{table:sn_lcdust}, we infer that SN 2006ax has little host galaxy dust extinction with the most likely value being $A_V = 0.01$ mag.
However, it is uncertain enough that $A_V = 0.12$ still lies within 68\% highest posterior density (HPD) contour.  By contrast, the $A_H$ estimate is near zero, and the 68\% contour lies within $A_H < 0.03$.  Even SN with low extinction benefit from observations in the $H$-band by reducing the uncertainty in the dust estimate.  Table \ref{table:sn_lcdust} lists summary statistics of the marginal posterior distribution of each host galaxy dust parameter for each SN, obtained from the MCMC samples.

\subsection{Intrinsic Correlation Structure of SN Ia Light curves in the Optical-NIR}

We use the hierarchical model to infer the intrinsic correlation structure of the absolute SN Ia light curves.  This correlation structure captures the statistical relationships between peak absolute magnitudes and decline rates of light curves in multiple filters at different wavelengths and phases.  We summarize inferences about light curve shape and luminosity across the optical and near infrared filters; a more detailed analysis of the intrinsic correlation structure of colors, luminosities and light curve shapes will be presented elsewhere.

\subsubsection{Intrinsic Scatter Plots}\label{sec:intrscatter}

\begin{figure}[t]
\centering
\includegraphics[angle=0,scale=0.4]{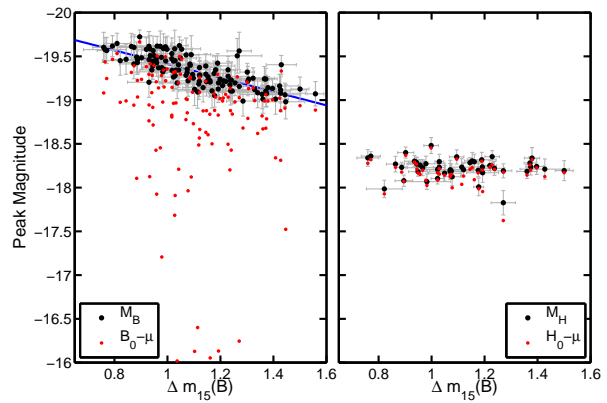}
\caption{\label{fig:MBH_d} (left) Post-maximum optical decline rate $\dmB$ versus posterior estimates of the inferred optical absolute magnitudes $M_B$ (black points) and the extinguished magnitudes $B_0 - \mu$ (red points).  Each black point maps to a red point through optical dust extinction in the host galaxy. The intrinsic light curve width-luminosity Phillips relation is reflected in the trend of the black points, indicating that SN brighter in $B$ have slower decline rates.  The blue line is the linear trend of \citet{phillips99}. (right) Inferred absolute magnitudes and extinguished magnitudes in the near infrared $H$-band.  The extinction correction, depicted by the difference between red and black points, is much smaller in $H$ than in $B$.  The absolute magnitudes $M_H$ have no correlation with the $\dmB$.  The standard deviation of peak absolute magnitudes is also much smaller for $M_H$ compared to $M_B$.}
\end{figure}

The hierarchical model fits the individual light curves with the differential decline rates model and infers the absolute magnitudes in multiple passbands, corrected for host galaxy dust extinction.   For each individual SN light curve, we can use the inferred local decline rates $\bm{d}^F$ to compute the $\Delta m_{15}(F)$ of the light curve in each filter.   In the left panel of Figure \ref{fig:MBH_d}, we plot the posterior estimate of the peak absolute magnitude $M_B$ versus its canonical $\Delta m_{15}(B)$ decline rate with black points.   The error bars reflect measurement errors and the marginal uncertainties from the distance and inferred dust extinction.   This set of points describes the well-known intrinsic light curve decline rate versus luminosity relationship \citep{phillips93}.   We also show the mean linear relation between $M_B$ and $\dmB$ found by \citet{phillips99}, who analyzed a smaller sample of SN Ia.  The statistical trend found by our model is consistent with that analysis.  The red points are simply the peak apparent magnitudes minus the distance moduli, $B_0 - \mu$, which are the extinguished peak absolute magnitudes $M_B + A_B$.    Whereas the range of extinguished magnitudes spans $\sim 3$ magnitudes, the intrinsic absolute magnitudes lie along a narrow, roughly linear trend with $\dmB$.  

In the right panel, we plot the intrinsic and extinguished absolute magnitudes of SN Ia in the $H$-band.  In contrast to the left panel, the differences between the intrinsic absolute magnitudes and the extinguished magnitudes are nearly negligible.   Notably, there is no correlation between the intrinsic $M_H$ in the NIR and optical $\dmB$.   This was noted previously by \citet{krisciunas04a} and \citet{wood-vasey08}.
The standard deviation of absolute magnitudes is much smaller in $H$ than in $B$, demonstrating that the NIR SN Ia light curves are good standard candles \citep{krisciunas04a,krisciunas04c,wood-vasey08,mandel09}.  Theoretical models of \citet{kasen06} indicate that NIR peak absolute magnitudes have relatively weak sensitivity to the input progenitor $^{56}\text{Ni}$ mass, with a dispersion of $\sim 0.2$ mag in $J$ and $K$, and $\sim 0.1$ mag in $H$ over models ranging from 0.4 to 0.9 solar masses of $^{56}\text{Ni}$.  The physical explanation may be traced to the ionization evolution of the iron group elements in the SN atmosphere. 

These scatter plots convey some aspects of the population correlation structure of optical and near infrared light curves that is captured by the hierarchical model.  In the next section, we further discuss the multi-band luminosity and light curve shape correlation structure in terms of the estimated correlation matrices.

\begin{figure}[t]
\centering
\includegraphics[angle=0,scale=0.4]{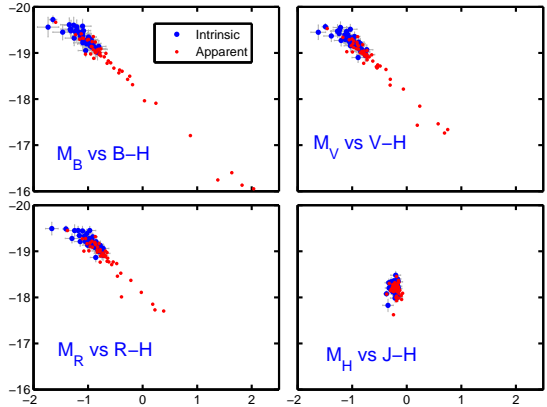}
\caption{\label{fig:M_color} Inferred absolute magnitudes $M_F$ (blue points) and the extinguished magnitudes $F_0 - \mu$ (red points) versus colors relative to NIR $H$ (intrinsic: blue points; apparent: red points).  Only SN with complete $BVRIJH$ data are plotted.  The intrinsically optically bright SN tend to be intrinsically bluer in optical-NIR color.  The $H$-band absolute magnitudes have no trend with intrinsic $J-H$ colors, which have a comparatively narrow distribution.  Note that the magnitude and color axes have the same scale in each panel.}
\end{figure}

Figure \ref{fig:M_color}  shows scatter plots of optical-near infrared colors $(B-H, V-H, R-H, J-H)$ versus absolute magnitude $(M_B, M_V, M_R, M_H)$ at peak.  The blue points are the posterior estimates of the inferred peak intrinsic colors and absolute magnitudes of the SN, along with their marginal uncertainties.  Red points are the peak apparent colors and extinguished absolute magnitudes, including host galaxy dust extinction and reddening.  These plots show correlations between the peak optical-near infrared colors and peak optical luminosity, in the direction of intrinsically brighter SN having bluer peak colors.  In contrast, the intrinsic $J-H$ colors have a relatively narrow distribution, and the near infrared absolute magnitude $M_H$ is uncorrelated with intrinsic $J-H$ color.

\subsubsection{Intrinsic Correlation Matrices}\label{sec:intrmatrices}

Using the hierarchical model, we compute posterior inferences of the population correlations between the different components of the absolute light curves of SN Ia.  This includes population correlations between peak absolute magnitudes in different filters, $\rho(M_F, M_{F'})$, correlations between the peak absolute magnitudes and light curve shape parameters (differential decline rates) in different filters, $\rho(M_F, \bm{d}^{F'})$, and the correlations between light curve shape parameters in different filters, $\rho(\bm{d}^F, \bm{d}^{F'})$.  They also imply correlations between these quantities and intrinsic colors.
This information and its uncertainty is captured in the posterior inference of the population covariance matrix $\bm{\Sigma}_\psi$ of the absolute light curve parameters $\{ \bm{\psi}_s \}$.  The posterior estimate of the absolute light curve population integrates over the posterior uncertainties in the individual light curves and the host galaxy dust estimates.

\begin{figure}[t]
\centering
\includegraphics[angle=0,scale=0.4]{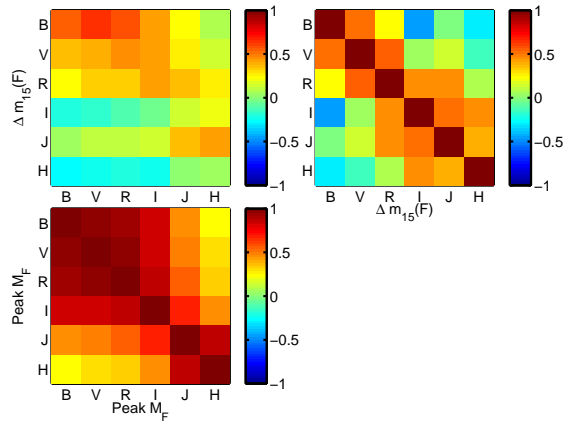}
\caption{\label{fig:corr_MM}The intrinsic correlation structure of optical-near infrared peak absolute magnitudes $M_F$ and decline rates $\dm(F)$.   The color scale encodes the strength of the correlation between any two intrinsic quantities.  The marginal mode for each correlation coefficient is depicted, integrating over the uncertainties of individual SN.
The posterior uncertainties in the correlations are computed but not shown.  (top left)  Correlations between peak intrinsic absolute magnitude and decline rate.  (top right) Correlations between decline rates for each pair of passbands.  (bottom)  Correlations between peak intrinsic absolute magnitudes for each pair of filters.}
 \end{figure}

In Figure \ref{fig:corr_MM}, we have distilled some of the information in this intrinsic covariance matrix to show the inferred intrinsic correlations.  For brevity, instead of depicting correlations with every differential decline rate $d^F_t$, we only show correlations with the canonical 15-day post-maximum decline rate in each filter, $\Delta m_{15}(F)$.   The correlations range from -1 to 1 and are color coded according to strength.  The joint uncertainties of the correlations are computed but not shown.
The bottom matrix shows the posterior inferences of the correlation matrix of peak absolute magnitudes.
 The optical luminosities and light curve shapes are strongly correlated with each other, but not with the NIR.    
 The $J$ and $H$ luminosities are strongly correlated with each other, but not with the optical.
Since the NIR luminosities have low intrinsic correlation with the optical luminosities, they provide independent information on the distance.  

The top left matrix shows the posterior inferences of correlation matrix of the $\Delta m_{15}$ decline rates and the peak absolute magnitudes in each filter.  The decline rates $\dm$ in $BVR$ exhibit strong correlations with peak optical absolute magnitudes, but they show low correlation with peak NIR absolute magnitudes in $J$ and $H$.  The decline rates $\dm$ in $IJH$ exhibit little correlation with luminosities in any of the optical or near infrared filters.

The top right matrix shows the posterior estimates of the correlations between the $\Delta m_{15}$ light curve decline rates in each filter.  The correlation matrix exhibits a band structure, with the largest correlations  neighboring the diagonal.   The decline rate $\dm$  in a particular filter is typically most strongly correlated with the $\dm$ in filters at  neighboring wavelengths.   This can be seen by examining each row of the correlation matrix.  The optical decline rates in $B$ and $V$ are strongly or moderately correlated with each other, but have low correlation with NIR decline rates.  Similarly, the decline rates in $IJH$ show strong or moderate correlation with each other, but have lower correlation with the decline rates in $B, V$.  The lack of strong correlation across the whole matrix indicates that the light curve shapes across optical and NIR wavelengths are unlikely to be adequately modeled with one degree of freedom.

These matrix plots depict some of the salient population correlation information of the SN Ia absolute light curves captured by the hierarchical model.  This inferred correlation structure  is used by the model to estimate luminosities from the light curves and to make distance predictions.

\subsection{Posterior Inference of the Host Galaxy Dust Population}\label{sec:postdust}

\subsubsection{Linear Correlation Dust Population Model}

In this section, we describe posterior inferences for the host galaxy dust population.  From the samples of the global posterior, Eq. \ref{eqn:globalposterior}, we can estimate the host galaxy dust extinction, $A_V$, and the slope of the extinction law, $r_V = 1/R_V$, for each object and their uncertainties from their marginal distributions.   We also estimate the characteristics of the dust population through the hyperparameters, $\tau_A, \bm{\beta}$, and  $\sigma^2_r$, while accounting for global uncertainties.

\begin{figure}[t]
\centering
\includegraphics[angle=0,scale=0.45]{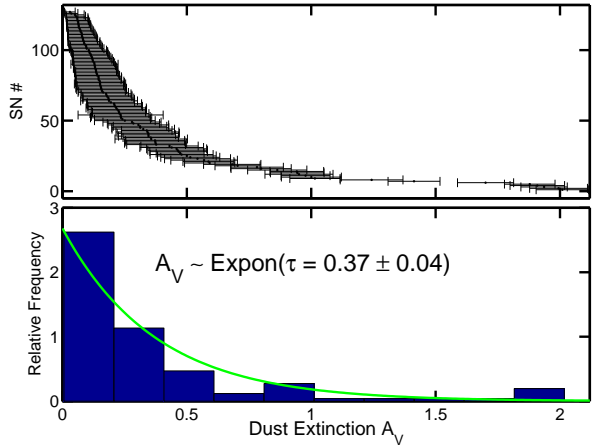}
\caption{\label{fig:Av_distr}The distribution of inferred host galaxy dust extinction $A_V$.  The hierarchical model estimates the extinction to each SN using the optical and near infrared light curves, and models the dust population. (top)  The $A_V$ estimates and uncertainties of each SN ranked from highest to lowest extinction. (bottom) The histogram of the modal $A_V$ estimates plotted against the fitted exponential distribution for the dust population.}
\end{figure}

In Figure \ref{fig:Av_distr}, we display the histogram of the $A_V$ estimates for each SN in the sample, along with the individual marginal estimates and their uncertainties.  This is compared against an exponential probability distribution with the marginal estimate of the extinction scale $\tau_A = 0.37 \pm 0.04$.   There appears to be an overabundance of object at $A_V > 1.5$ compared to the exponential distribution.  We explore this further in \S \ref{sec:checks}.

\begin{figure}[b]
\centering
\includegraphics[angle=0,scale=0.45]{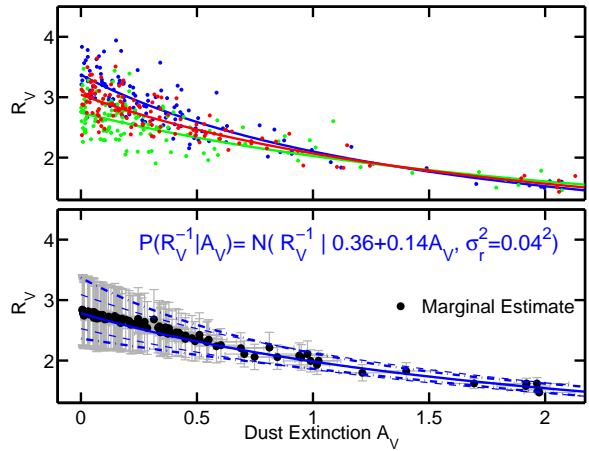}
\caption{\label{fig:dustpop}Apparent correlation between host galaxy dust visual extinction $A_V$ and the dust law slope $R_V$ in the sample of SN Ia.   This model assumes a dust population where $R_V^{-1}$ has a linear trend with $A_V$ with some rms scatter $\sigma_r$.   The linear regression coefficients and residual scatter ($\bm{\beta}, \sigma^2_r$) are estimated from the marginal global posterior distribution.  (top)  The points of each color and the regression relation are different probable realizations of the ($A_V, R_V$) for each SN and dust population hyperparameters,  $\bm{\beta}$ and $\sigma_r^2$, obtained from snapshots of the MCMC .  The $R_V$ estimates at low extinctions have more uncertainty than those at high extinction, as reflected by the scatter of points with different colors. (bottom) Averaging over all probable realizations, we plot the \emph{inferred} marginal posterior mode of $(A_V, R_V)$ and their marginal uncertainties for each SN with the marginal estimates of the regression model.  When the individual $R_V$ estimates for single SN are very uncertain, they tend to be pulled toward the population mean value (for its extinction $A_V$) using partial pooling.
The data favor an apparent non-zero correlation between $A_V$ and the dust slope $R_V^{-1}$.   SN Ia light curves with low to moderate extinction are consistent with the Milky Way average $R_V \sim 3.1$ for interstellar extinction, but for highly extinguished SN, a low value of $R_V \lesssim 2$ is favored. }
\end{figure}

In Figure \ref{fig:dustpop}, we display the estimates of $(A_V, R_V)$ for the $m = 1$ population model, described in \S \ref{sec:dustpop}. This model \emph{assumes} that the mean trend of $r_V$ versus $A_V$ is linear in $A_V$.   Fitting the hierarchical model then entails computing posterior estimates of $(A_V, r_V)$ for individual objects and the population trend, parameterized by $\bm{\beta}$, $\sigma_r^2$.  
For SN at low $A_V$, the $r_V$ parameter for each individual SN cannot be estimated precisely, since it only enters into the extinction model, Eq. \ref{eqn:AFt}, and thus, the likelihood, multiplied by $A_V$.  For these SN, there is not enough information in individual light curves to distinguish between the individual $r_V$ estimates, and so the model pools them towards the group mean or trend.  At  high $A_V$, the $r_V$ parameter can be estimated more precisely for each SN, so they can be individually distinguished.  In the top panel, we show the $A_V, R_V$ values for each SN for three joint samples from the MCMC chain.  A joint sample of $\{A_V^s, R_V^s\}, \bm{\beta}, \sigma^2_r$ represents a single probable realization of these parameters given the data, and  is labelled by a single color.   The $R_V$ estimates at low $A_V$ show considerable scatter  between samples, reflecting the underlying uncertainty.  At high $A_V$, there is less scatter between individual SN and between samples, reflecting the increased precision for estimating $R_V$.  In the bottom panel,  each point and error bar represents the marginal estimate, averaging over all the MCMC samples, of $(A_V, R_V)$ for each SN.

\begin{figure}[t]
\centering
\includegraphics[angle=0,scale=0.45]{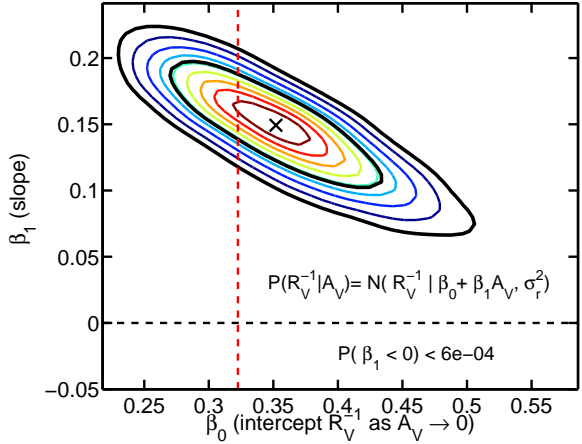}
\caption{\label{fig:post_betas}Marginal posterior distribution of the linear regression coefficients of the dust population model assuming a linear mean trend between $R_V^{-1}$ and $A_V$.  The parameter $\beta_1$ is the slope of $R_V^{-1}$ against extinction $A_V$, and $\beta_0$ is the  population mean value of $R_V^{-1}$ in the low extinction limit.  The two-dimensional mode is marked, and the inner and outer solid black lines contain 68\% and 95\% of the marginal probability, respectively.
The posterior estimate of $\beta_0$ is consistent with $R_V = 3.1$ (vertical red dashed line) and inconsistent with $R_V < 2$, $P(\beta_0 > 0.5) = 0.02$.  The posterior estimate of the regression slope $\beta_1$ is extremely inconsistent with zero (horizontal dashed line).}
\end{figure}

In Figure \ref{fig:post_betas},  we show the bivariate marginal probability density of the regression parameters $\bm{\beta} = (\beta_0, \beta_1)$, obtained from the MCMC samples.  The joint mode, and the 68\% and 95\% highest posterior density contours are shown.  The intercept $\beta_0$ represents the population mean value of $r_V$ at vanishing $A_V \rightarrow 0$, and $\beta_1$ represents the population mean linear trend of $r_V$ against $A_V$.   Also shown is the value of $r_V$ corresponding to the Milky Way interstellar average $R_V = 3.1$.   The intercept $\beta_0$ at vanishing $A_V$ is uncertain, but consistent with the Milky Way average within $1 \sigma$.  The regression slope $\beta_1$ is positive with zero excluded from the 95\% credible region.  The marginal estimates of each of the regression parameters are listed in Table \ref{table:Rv}.
The marginal estimate of  $\beta_0 = 0.35 \pm 0.05$ can be compared against $r_V = 0.32$ for the Milky Way average.  The characteristic value of $R_V$ as $A_V \rightarrow 0$, $\beta_0^{-1}$, is uncertain because of the difficulty of determining $R_V$ for low-extinction objects.   The marginal posterior density of $\beta_0^{-1}$ has a non-gaussian profile: the mean is 2.9, the mode is 2.7, and the interval containing 68\%  of the highest probability density is $[2.3, 3.3]$.  The marginal probability that $\beta_0^{-1} < 2$ is $p = 0.02$.
The marginal estimate of the slope is  $\beta_1 = 0.15 \pm 0.03$. This is a strong indication of a differential trend of $r_V$ vs. $A_V$ in the host galaxy dust population of nearby SN.   

\input{Rv_scenarios.tex}

These results were consistent when changing the peculiar velocity dispersion $\sigma_\text{pec}$ from 150 to 300 $\text{ km s}^{-1}$.  The posterior mean of $\beta_0^{-1}$ was $2.8$, the mode was $2.5$, and the $68\%$ interval containing highest probability density was $[2.1, 3.3]$. The marginal estimate of the slope is $\beta_1 = 0.14 \pm 0.04$.

\begin{figure}[b]
\centering
\includegraphics[angle=0,scale=0.45]{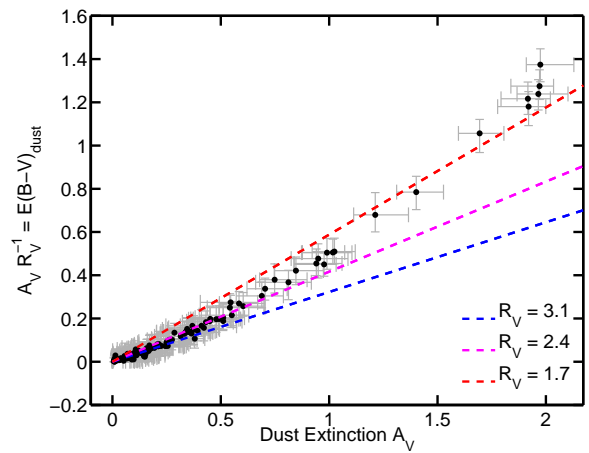}
\caption{ \label{fig:ebv_vs_Av} Marginal posterior estimates of inferred color excess $E(B-V)$ due to host galaxy dust versus inferred extinction $A_V$, assuming the linear correlation model.  This model assumes a dust population where $R_V^{-1}$ has a linear trend with $A_V$ with some rms scatter $\sigma_r$ The SN Ia at lower extinction ($A_V \lesssim 1$) have implied color excesses consistent with $R_V = 2$ to $3$.  The SN Ia at higher extinction favor a dust law with $R_V < 2$.}
\end{figure}

Figure \ref{fig:ebv_vs_Av} plots the posterior estimate of the inferred optical reddening $E(B-V) \equiv A_B - A_V$ due to host galaxy dust versus the estimated dust extinction $A_V$, assuming the linear correlation model.  The reddening estimates at $A_V > 1.5$ favor a $R_V = 1.7$ reddening law, whereas at lower extinction, $A_V < 1$, the reddening estimates are consistent with $2.4 \lesssim R_V \lesssim 3.1$.   

\subsubsection{Step Function Dust Population Model}

The linear correlation model, $m=1$, assumes that the mean trend of $r_V$ with $A_V$ is linear across the entire range of $A_V$.   However, we do not know if this assumption is true.  To test the sensitivity of the apparent differential trend in $r_V$ vs. $A_V$ to the linear correlation assumption of the $m=1$ model, we fit alternate models using the ``Step'' Case 6 of \S \ref{sec:dustpop}.  Instead of using all the SN over the range of $A_V$ to determine a linear correlation in the host galaxy dust population, this case groups together only SN in the same interval of $A_V$ to determine their group mean and variance of $r_V = R_V^{-1}$ in each bin.  We have re-trained the hierarchical model using the step function assumptions, first by dividing the range in $A_V$ into ``high'' ($A_V > 0.8$) and ``low'' ($A_V < 0.8$) extinction bins.  Second, we subdivided those bins and re-trained the model using four bins in extinction, $A_V$.

Marginal posterior estimates for the ``2-step'' model are listed in Table \ref{table:Rv_2steps}.  The low extinction bin, with $A_V < 0.8$, has a group $R_V$ mean of $2.3 \pm 0.3$, while the high extinction bin, $A_V > 0.8$ has a group $R_V$ mean of $1.7 \pm 0.1$.   We examined the marginal posterior probability of the difference between the group means $\mu_r$ at low and high extinction.  This calculation takes into account the posterior covariance between the estimates and marginalizes over uncertainties in the other parameters.    The tail probability that $\mu_r$ of the low extinction bin is greater than $\mu_r$ of the high extinction bin is denoted $p_\text{tail}$, and is computed directly from the MCMC samples.  We find less than 1\% probability that the difference is positive, suggesting that the difference is significant.

\input{Rv_steps2.tex}

For the ``4-step'' model, the posterior inferences of the hyperparameters in each of the four intervals in $A_V$ are listed in Table \ref{table:Rv_4steps}.  The group means, $\mu_r$, for each interval display the same trend of lower $R_V$ for higher $A_V$.  The bin with $0 \le A_V < 0.4$ has a group mean consistent with $R_V \approx 3$, the interstellar average for the Milky Way.    However, the group mean for the lowest extinction SN is uncertain due to the difficult of determining $r_V$ at low $A_V$.  The marginal posterior density of the chracteristic $R_V$ in the lowest extinction bin, $\mu_r^{-1}$, is non-gaussian:  the mean is 2.9, the mode is 2.5 and the interval containing 68\% of the highest probability density is $[2.1, 3.3]$.  The marginal probability that $\mu_r^{-1}$ of the lowest extinction bin is less than 2 is $p = 0.04$.
The highest extinction bin, $A_V > 1.25$, favors a group mean $R_V = 1.6 \pm 0.1$.   We calculated $p_\text{tail}$, the marginal probability that the group mean $\mu_r$ of each bin is greater than $\mu_r$ of the highest extinction bin.  Each of the lower extinction bins had significantly different group mean $\mu_r$ estimates than that of the highest extinction bin.

\input{Rv_steps4.tex}

For an assumed  peculiar velocity dispersion $\sigma_\text{pec} = 300 \text{ km s}^{-1}$, the results were consistent.  For the lowest extinction bin, the non-gaussian marginal probability density of $\mu_r^{-1}$ had a peak at 2.5, with mean 2.7 and a 68\% interval $[2,3]$.  The posterior mean and standard deviation of $\mu_r^{-1}$ in the highest extinction bin was $1.5 \pm 0.1$.  The inferences for the alternate step function model confirm the trend seen in the $m=1$ linear correlation model.  We discuss the implications of these findings in \S \ref{discussion}.

\subsubsection{Other Dust Population Models}

Posterior inferences of the hyperparameters using the other models for the host galaxy dust population (\S \ref{sec:dustpop}) are listed in Table \ref{table:Rv}.   In the case of complete pooling (CP), in which it is assumed that all SN have the same value of $R_V$, the marginal estimate of that value is $R_V = 1.6 \pm 0.1$.  In the population model $m = 0$,  in which each $r_V$ is drawn from a Gaussian with mean independent of $A_V$, we find a population mean with a similar value ($\mu_r^{-1} = 1.7 \pm 0.1$).  These results indicate that  the highly extinguished SN dominate the estimate of the global constant or population mean in these cases, since their individual $R_V$ estimates are the most precise.  The CP and $m = 0$  are special cases of the  $m=1$ model.  If  the $m=0$ model were favored then when fitting the $m=1$ model we should have found $\beta_1 \approx 0$.  If CP were favored then we would have also found that $\sigma^2_r = 0$.   We inferred none of those in the expanded $m=1$ model; this illustrates the pitfalls of those simpler assumptions.

\section{Model Checks}\label{sec:checks}

After fitting the hierarchical model by computing the global posterior density, Eq. \ref{eqn:globalposterior}, using our \textsc{BayeSN} code, we checked the model fit using several methods.    We did this to ensure first that the MCMC code was fitting the assumed statistical model to the data set, and to diagnose technical or algorithmic errors.  Secondly, we checked the fit of the hierarchical model to the observed sample to look for disagreements between the assumptions and the observed data.  Third, we tested the robustness of the model to the training set and evaluated distance prediction error by performing extensive cross-validation (\S \ref{sec:cv}).  
 For individual SN, we inspected the fits of the light curve model to the photometric data (\S \ref{sec:individuals}).   
 
\begin{figure}[t]
\centering
\includegraphics[angle=0,scale=0.45]{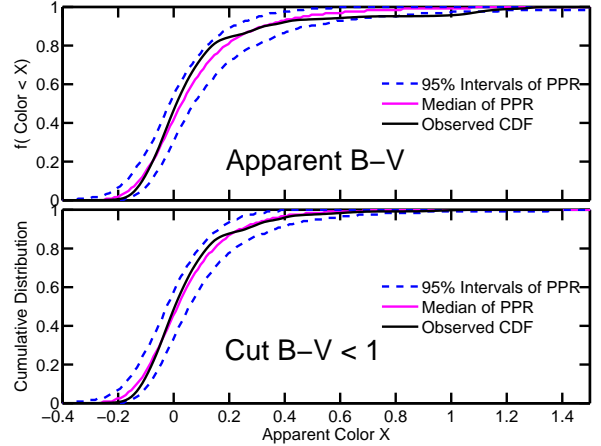}
\caption{\label{fig:ppr_appBV} Comparison of the cumulative distribution of peak apparent $B-V$ colors to those of posterior predictive replication sets randomly generated from the trained model.  (top)  When the hierarchical model is trained on our full sample, the distribution of apparent colors of the observed sample has a thicker tail towards the extreme red ($B-V \gtrsim 0.8-1$ mag) than most of the the replicated sets.  (bottom)  When the hierarchical model is retrained on a sample restricted to  apparent $B-V < 1$, the apparent color distribution agrees well with replicated SN sets.  This suggests that the exponential model population distribution for extinction inadequately accounts for the number of SN observed at very high reddening.}
\end{figure}

To check the fit of the model population distributions to the apparent distributions of the data set, we performed posterior predictive model checks \citep{rubin84,gelman96,gelman_bda}.    From the trained hierarchical model, we generated a new random set of apparent light curves of the same size as the observed sample.   The replicated set was generated by sampling forward through the directed acyclic graph, Fig. \ref{fig:dag}.  The distribution of apparent properties of the replicated light curves was compared to those of the observed set.  We illustrate such a comparison of peak apparent $B-V$ optical colors in Figure \ref{fig:ppr_appBV}.  We generated 1000 replications, each containing the same number of SN as the observed sample.  The apparent colors of the SN within each replicated set have a cumulative distribution function.  The distribution of apparent $B-V$ colors is the convolution of the intrinsic $B-V$ color distribution and the dust $E(B-V)$ color excess distribution implied by the extinction distribution.
The set of replications form an ensemble of color distributions, reflecting random sampling variation and posterior uncertainty in the model.  For each value of the $B-V$ color we show the median, 2.5\% and 97.5\% quantiles of the ensemble of CDFs at that value.    The black curve is the CDF of the apparent colors of the observed data set.  If the observed CDF lies outside the 95\% range of the replications, then the observed distribution disagrees with the model's replications.

In the top panel of Fig. \ref{fig:ppr_appBV}, the observed distribution of peak apparent $B-V$ colors has a significantly thicker tail towards redder (positive) colors than the replicated distributions.  This suggests that the number of very red SN with $B-V > 1$ is large compared to what can be expected with the exponential model for the dust population.   The abundance of very red SN in the nearby sample might be a consequence of preferred selection of these events for follow-up observation.

To test whether the model distribution adequately describes the SN with less reddening , we removed the 4\% of SN with apparent $B-V > 1$ from the training set, retrained the whole hierarchical model, and again generated posterior predictive replciations (bottom panel, Fig. \ref{fig:ppr_appBV}).   There is good agreement between apparent color distributions of the ensemble of replications and the observed data set.    With the color cut, the estimated exponential scale of the  extinction distribution decreased from $\tau_A = 0.37 \pm 0.04$ mag to $\tau_A = 0.28 \pm 0.04$, so that the model captures a dust extinction distribution with a thinner tail, which implies a narrower apparent color distribution.   We found that $\bm{\beta}$ was consistent within the uncertainties with the values found by using the whole sample.  This demonstrates that the trend is not determined just by the reddest outliers of the SN sample.

\begin{figure}[b]
\centering
\includegraphics[angle=0,scale=0.45]{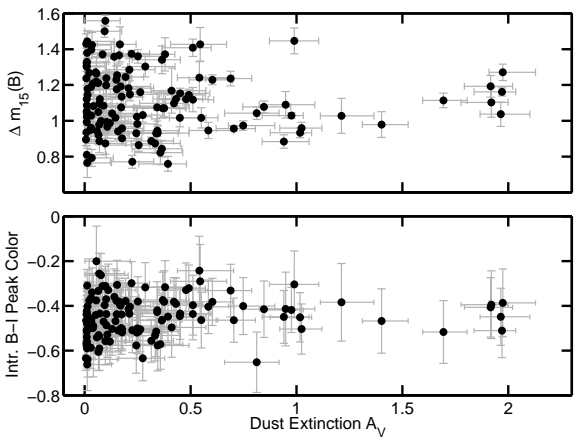}
\caption{\label{fig:dec_vs_dust} (top) The fitted intrinsic $\dmB$ decline rate of SN Ia light curves  versus inferred dust extinction.    There is no significant trend between $\dmB$ and $A_V$.  (bottom) The inferred intrinsic $B-I$ color at peak versus the inferred dust extinction.   There is no significant trend between peak intrinsic $B-I$ and $A_V$.}
\end{figure}

A key assumption of the model is that the two populations, the SN Ia light curves and the dust extinction, are statistically independent.   This entails that an intrinsically faint or red supernova has the same chance of encountering a particular level of host galaxy dust extinction as an intrinsically bright or blue supernova.
We expect that  the amount of extinction to SN should be uncorrelated with the intrinsic properties of the SN Ia light curves.    A significantly non-zero relationship between the two might indicate a miscalibration of the model, possibly related to a confusion between intrinsic color variation and dust extinction.    We tested this hypothesis, as shown in Figure \ref{fig:dec_vs_dust}, where we plot the fitted intrinsic $\dmB$ decline rates and the inferred intrinsic $B-I$ color of SN Ia light curves versus the inferred dust extinction.
The plots shows the expected lack of correlation between the parameters from the two populations, and is a consistency check on the model fit.

\section{Distance Prediction}\label{sec:prediction}

\subsection{Hubble Residuals under Resubstitution}

After training the model on all the SN in the sample ($\mathcal{D}, \mathcal{Z}$), the distance modulus for each SN can be estimated by re-substituting each light curve set into the model, and computing the posterior predictive density $P( \tilde{\mu}_s | \, \tilde{\mathcal{D}}_s, \tilde{z}_s; \mathcal{D}, \mathcal{Z})$, which marginalizes over the uncertainty in the trained model.  The expected value of this density is $\mu_{\text{resub}}^s$.  The Hubble residual is the difference between the resubstitution distance modulus and the distance modulus expected from the redshift and the Hubble law, $f(z_s)= \mathbb{E}(\mu_s | z_s) $.  The estimates $ \mathbb{E}(\mu_s | z_s) $, $\sigma_{\mu,s}$ and $\mu_\text{resub}$ are listed in Table \ref{table:mu}.
The uncertainty-weighted mean square resubstitution error, $\text{err}^2_\text{resub}$  is computed as a sum over all SN, using Eq. 31 of \citet{mandel09}.   For the $m=1$ dust model, the error-weighted \textit{rms} of the Hubble residuals at $cz > 3000 \text{ km s}^{-1}$ is $0.13$ mag for the full sample.   However, for the SN with NIR data,  the resubstitution error at $cz > 3000 \text{ km s}^{-1}$ is $0.10$ mag, and those with only optical data have a resubstitution error of 0.14 mag.

\subsection{Cross-Validation \& Prediction Error}\label{sec:cv}

For finite samples, the rms Hubble diagram residual of the training set SN is an optimistic estimate of the ability of the  statistical model  to make accurate distance predictions given the supernova observables.   This is because it uses the supernova data twice: first for estimating the model parameters (training), and second for evaluating the residual error.   To evaluate predictive performance and guard against over-fitting with a statistical model based on finite data, we should estimate the prediction error for SN not included in the training set (``out-of-sample'').  We use cross-validation (CV) to evaluate the utility of optical and NIR light curves for accurately predicting distances in the Hubble diagram, and to test the sensitivity of the model to the finite training set.    The importance of cross-validating statistical models for predicting SN Ia distances has been discussed by \citet{mandel09} and \citet*{blondin11}.

To estimate the distance prediction error of our statistical model, we have performed bootstrap cross-validation.  This method was first used for assessing distance predictions of SN Ia light curve models by \citet{mandel09}.  
From the full set of SN, a new bootstrapped training set is created by sampling with replacement individual SN up to the same size as the original set.  The complement of this new training set forms a validation or prediction set.  The training set light curves and redshifts are used to build the statistical model for SN Ia light curves, with the model hyperparameters estimated using hierarchical Bayesian inference and MCMC.  The prediction set light curves are used to generate distance predictions for those SN, which are then compared to the Hubble distances expected from their redshifts.

We randomly  bootstrapped 30 training sets, so that on average each SN was held out for distance prediction 11 times.
For each SN $s$, the expected value of the posterior predictive probability density, $\mu_{\text{pred},B}^s \equiv \mathbb{E}(\tilde{\mu}_s | \tilde{\mathcal{D}}_s, \tilde{z}_s ; \mathcal{D}^B, \mathcal{Z}^B)$, is a point estimate of the distance modulus prediction under the training set data $\mathcal{D}^B, \mathcal{Z}^B$ for training set $B$.
The .632 bootstrap estimate  \citep{efron83,efron97} of rms prediction error is computed using the sum of uncertainty-weighted squared prediction errors over all bootstrapped sets, as described by Equations 32 and 33 of \citet{mandel09}.
For the $m=1$ dust model, we list in Table \ref{table:mu} we list the predicted distance modulus for each SN, averaged over all the training sets $B$ that do not include that SN, as $\bar{\mu}_\text{pred}$.    The standard deviation of predictive uncertainty, i.e. the square root of $\text{Var}[\tilde{\mu}_s | \, \tilde{\mathcal{D}}_s, \tilde{z}_s;  \mathcal{D}^B, \mathcal{Z}^B]$, averaged over the training sets $B$ not containing that SN, is $\sigma_\text{pred}$.   This measures the precision with which the trained model makes a distance prediction for a particular SN.  The standard deviation of the point estimates $\mu_{\text{pred},B}^s$ over all of those training sets, $s_\text{pred}$, is a measure of the sensitivity of the predicted distances to resampling the training set.

\begin{figure}[t]
\centering
\includegraphics[angle=0,scale=0.38]{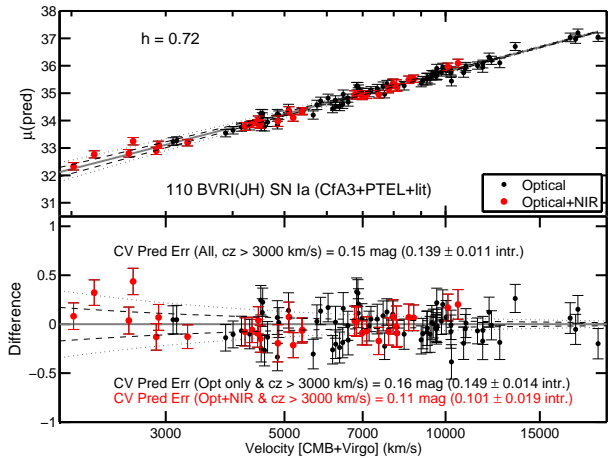}
\caption{\label{fig:hubble_cv_cfa}Cross-validated Hubble diagram computed with \textsc{BayeSN} for the low-$z$ nearby set of CfA and literature SN.   Red points indicate the SN with joint optical $BVRI$ and near infrared $JH$ data.  Black points have only optical data.   The dashed (dotted) line indicates the magnitude uncertainty in $\mu(z)$ for $\sigma_\text{pec} =$ 150 (300) $\text{ km s}^{-1}$.  We perform cross-validation with 30 bootstrapped training sets to estimate the out-of-sample prediction error and test the sensitivity of the model predictions to the finite sample.  The \emph{rms} prediction error in distance modulus for SN with optical  light curve data only at $cz > 3000 \text{ km s}^{-1}$ is 0.16 mag.  The SN with optical and near infrared light curve data have an \emph{rms} prediction error at $cz > 3000 \text{ km s}^{-1}$ of 0.11 mag.  The maximum likelihood estimate of the rms \emph{intrinsic} prediction error is shown in parentheses, assuming $\sigma_{\text{pec}} = 150$ km s$^{-1}$.  For a velocity dispersion $\sigma_{\text{pec}} = 300$ km s$^{-1}$, the total rms prediction error remains the same, but the scatter attributed to intrinsic prediction error is $0.129 \pm 0.016$ mag for SN with optical data only, or $0.081 \pm 0.026$ mag for SN with optical and NIR light curve data.
These results indicate that one can make more accurate distance predictions with SN Ia with combined optical and NIR data than with optical data alone.}
\end{figure}

Figure \ref{fig:hubble_cv_cfa} shows the predicted distances to the SN using bootstrap cross-validation.   
For Hubble flow SN at $cz > 3000 \text{ km s}^{-1}$, the cross-validated prediction error is 0.15 mag overall.  For the SN with optical and NIR data, the prediction error is estimated to be 0.11 mag, and for SN with optical light curves alone, the rms prediction error is 0.16 mag.  
The predicted distances to SN with optical and NIR light curve measurements have a smaller scatter in the Hubble diagram than those with only optical data.   These estimates of Hubble diagram scatter can be compared to the 0.18-0.22 mag rms found for the CfA3 sample using the MLCS2k2 and SALT2 methods \citep{hicken09b}.

The weighted rms  prediction error measures the total Hubble diagram scatter, comprised of at least two components:  a dispersion associated with unknown and random peculiar velocities with respect to the Hubble expansion, and an \emph{intrinsic} variance that represents a floor to the precision of distance predictions.
We compute this intrinsic component of the prediction error using the maximum likelihood estimator described in Appendix B of \citet*{blondin11}.
Assuming a velocity dispersion $\sigma_{\text{pec}} = 150$ km s$^{-1}$, the rms intrinsic prediction error was $0.15 \pm 0.01$ mag for SN with optical data only, and $0.10 \pm 0.02$ for SN with optical and near infrared light curves.  For $\sigma_{\text{pec}} = 300$ km s$^{-1}$, the weighted rms prediction error remains the same, but the scatter attributed to intrinsic prediction error is $0.13 \pm 0.02$ mag for SN with optical data only, or $0.08 \pm 0.03$ mag for SN with optical and NIR light curve data.  These estimates of intrinsic prediction error are smaller when a larger  $\sigma_{\text{pec}}$ is assumed because more of the Hubble diagram scatter is attributed to random galaxy motions.

\begin{figure}[t]
\centering
\includegraphics[angle=0,scale=0.38]{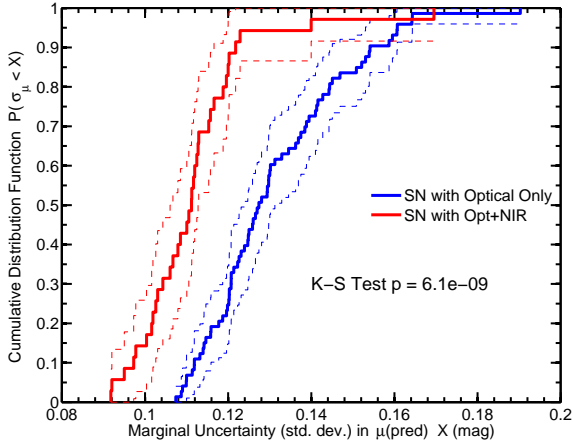}
\caption{\label{fig:mu_unc} Cumulative distributions of the marginal predictive uncertainties of distance moduli of SN Ia.  The uncertainty in the predicted distance modulus is represented by a probability density after integrating over the other uncertainties in the dust estimates, light curve fits and $K$-corrections, and the population.  We show the CDFs of the standard deviations of the predictive probability distributions of the individual SN distance moduli.  The SN with optical and NIR light curve measurements (red) typically have a smaller distance uncertainties (higher precision) than those with only optical light curve data (blue).   
The dashed lines represent 95\% confidence intervals of the respective CDFs.  The two distributions are highly discrepant according to the Kolmogorov-Smirnov test.    Using combined optical and NIR light curve data, the hierarchical model makes distance predictions with smaller estimated uncertainty and higher precision than it does with optical data alone.}
\end{figure}

The predictive variance $\sigma^2_{\text{pred}}$ measures the uncertainty with which the hierarchical model predicts the distance modulus of each individual SN, after marginalizing over the uncertainties in the training set, SN Ia and dust populations.   Figure \ref{fig:mu_unc} shows the distributions of predictive uncertainties (standard deviations) in the individual SN distance moduli predicted from this model.    The cumulative distribution of the predictive posterior standard deviations for SN with optical light curve data only is compared to that of SN with optical and NIR data.   For SN Ia with joint optical and NIR data, the predictive uncertainties are typically between 0.10 and 0.12 mag, whereas for SN Ia with optical data only, the uncertainties mostly lie between 0.12 and 0.16 mag.  A simple Kolmogorov-Smirnov test verifies that these precision distributions are inconsistent.
This demonstrates that the hierarchical model estimates the distances to SN Ia with optical and NIR light curves with smaller uncertainty than those of SN Ia with only optical data.  

The sample variance of the distance predictions for a single SN over bootstrapped training sets, $s^2_\text{pred}$ in Table \ref{table:mu}, is always much smaller than the uncertainty variance $\sigma^2_{\text{pred}}$ of a single prediction, and is smaller than the mean square error over the set of SN in the Hubble diagram.  The typical value of $s_\text{pred}$ over the set of SN is $\sim 0.03$ mag.  This demonstrates that our model's distance predictions to individual SN are fairly robust to perturbing the composition of the training set; the sensitivity of predictions to resampling is of order a few hundredths of a magnitude.    With a larger set of optical and NIR light curves, this sensitivity could be reduced further.

\begin{figure}[b]
\centering
\includegraphics[angle=0,scale=0.4]{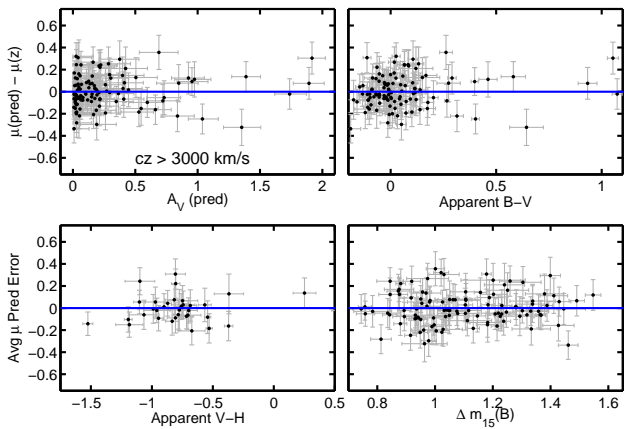}
\caption{\label{fig:pred_errs}Cross-validated distance prediction errors $\mu_\text{pred} - \mu(z)$ versus observed and inferred SN quantities of interest.  The distance modulus prediction errors of the model, averaged over 20 bootstrapped training sets, do not show statistically significant trends with respect to host galaxy dust extinction $A_V$, apparent optical color $B-V$, apparent optical-near infrared color $V-H$ or optical decline rate $\dmB$.  Fitted regressions have slopes consistent with zero (blue).}
\end{figure}

We examined the cross-validation prediction errors to check for systematic trends against observable or inferred quantities, as possible signs of model misfit.  In Figure \ref{fig:pred_errs} we show a scatter plot of the prediction error for each SN versus an observable or inferred quantity.  We find no significant trends of prediction error versus predicted dust extinction $A_V$, the apparent optical colors at peak (e.g. $B-V$), apparent optical-near infrared colors at peak (e.g. $V-H$), or optical light curve shape, summarized by the canonical $\dmB$.   Linear regressions fit to the prediction errors versus each quantity yield both slopes and intercepts that are statistically consistent with zero.

\subsection{Distance Error Comparison with CSP Light Curves}

\begin{figure}[t]
\centering
\includegraphics[angle=0,scale=0.4]{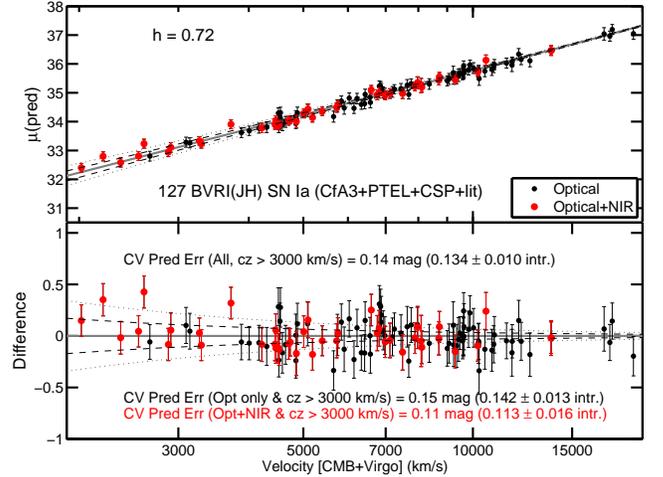}
\caption{\label{fig:hubble_cv_cfacsp}Cross-validated Hubble Diagram computed with \textsc{BayeSN} for the low-$z$ nearby training set using CfA, CSP and literature SN.  Red points indicate the SN with joint optical $BVRI$ and near infrared $JH$ data.  Black points are SN with only optical data.   The dashed (dotted) line indicates the magnitude uncertainty in $\mu(z)$ for $\sigma_\text{pec} =$ 150 (300) $\text{ km s}^{-1}$.
We perform bootstrap cross-validation to estimate the out-of-sample prediction error and test the sensitivity of the model predictions to the finite sample.  
The \emph{rms} prediction errors are consistent with those of Fig. \ref{fig:hubble_cv_cfa}. }
\end{figure}

We augmented our SN sample with 27 nearby SN recently published by the Carnegie Supernova Project \citep{contreras10}, and again performed the cross-validations to produce predictions for each SN.   There were 10 SN that were contained both in the CSP sample and the CfA3+PAIRITEL sample.  To avoid including duplicate light curves for the same SN in the joint sample, we selected the CSP light curves in those cases, since this resulted in the retention of the most optical and NIR data.   We recomputed training and prediction under the $m=1$ dust population model.  The Hubble diagram of these distance predictions is shown in Figure \ref{fig:hubble_cv_cfacsp}.  The results are consistent with the previous Hubble diagram: the total rms dispersion at $cz > 3000 \text{ km s}^{-1}$ was 0.15 for SN with optical data only, and 0.11 for SN with optical and near infrared light curves.
We compared the distribution of distance modulus errors $\bar{\mu}_\text{pred} - f(z)$ for Hubble flow SN with optical and NIR data from the CSP sample with that of the ``CfA+literature'' sample.  Using a two-sample Kolmogorov-Smirnov test we cannot rule out that they are from the same distribution ($p = 0.88$).   The distance predictions of each set are statistically consistent, so it is reasonable to analyze the combined set.

\subsection{Cross-validation with Different $R_V$ Assumptions}

In this section, we investigate the effect of different model assumptions about the dust population on distance predictions.   For each case of \S \ref{sec:dustpop}, we computed cross-validated distance predictions for SN in the Hubble flow at $cz > 3000 \, \text{km s}^{-1}$ by generating 20 bootstrapped sets for training and predicting the distance moduli for the complementary validation set.  For these computations, we used the ``CfA+CSP+literature'' sample of 127 SN Ia.  Table \ref{table:Rv} displays the results of these calculations, including the marginal posterior estimates of the hyperparameters in each case, and the 0.632 estimate of total prediction error for the SN with optical light curves only, and for the SN with both optical and near infrared data.

The case with fixed $R_V = 3.1$ (the Milky Way interstellar average) for all SN leads to the worst distance predictions (0.20 mag for optical, 0.13 mag for optical and near infrared).     The cases of complete pooling (all SN have $R_V$ with the same value) or partial pooling with $m = 0$ (the $R_V$s come from a population independent of the $A_V$ value), or even no pooling (each $R_V$ estimated independently for each SN), \emph{rms} prediction errors are about 0.15 mag to 0.16 mag for optical light curves only, and 0.12-0.13 mag for optical plus NIR light curves.

If we model a potential population correlation between $A_V$ and $R_V^{-1}$, using either linear or step function models, we find the smallest cross-validated distance prediction errors, both for SN with optical data only (0.15 mag), and for the SN with optical and NIR light curves (0.11 mag).  These are both significant improvements over the rms prediction errors under the assumption that $R_V = 3.1$ has the mean value for Milky Way interstellar dust, and are also marginally better than those of the other cases.  However, the statistical sampling uncertainty is about $\pm 0.01$ mag, so it is difficult to draw significant distinctions between the rms cross-validated prediction errors of the latter five cases.  An analysis of a larger, future sample of optical and NIR light curves will help to further discriminate between these competing cases.

It is notable that the change in the rms distance modulus prediction error for SN with optical light curves alone is 0.05 mag between the worst case and best case dust population models, whereas this change for SN with optical and NIR light curves is only 0.02 mag.  This highlights the advantage of including the NIR data; since the $H$-band provides a good standard candle, the model can rely mostly on the NIR light curves to provide distance estimates that are both less vulnerable to host galaxy dust, and less sensitive to the assumptions about the dust.

\subsection{Improving Constraints on Dust and Distance with Optical and NIR Data}\label{sec:improving}

\begin{figure}[t]
\centering
\includegraphics[angle=0,scale=0.225]{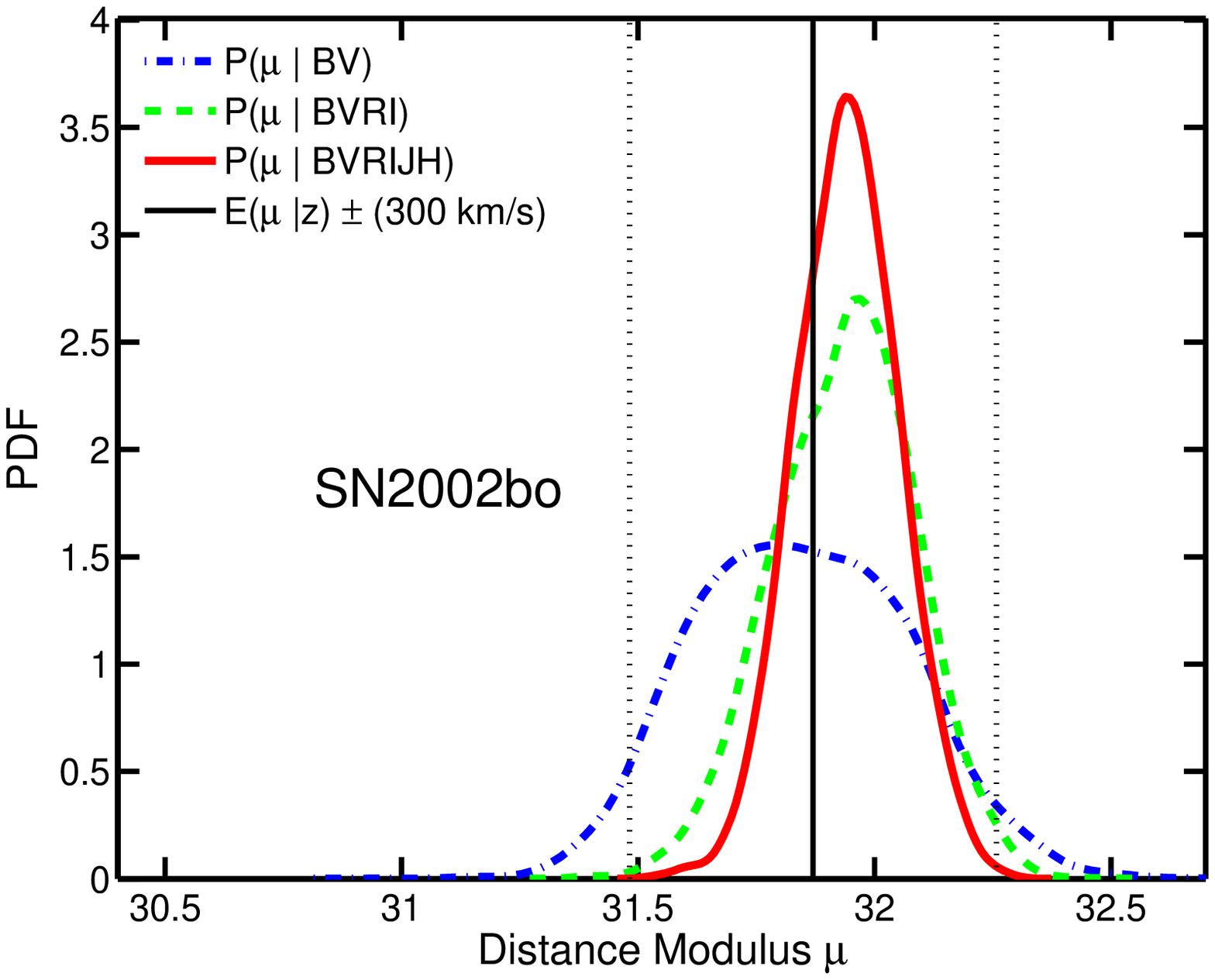}
\hfill
\includegraphics[angle=0,scale=0.225]{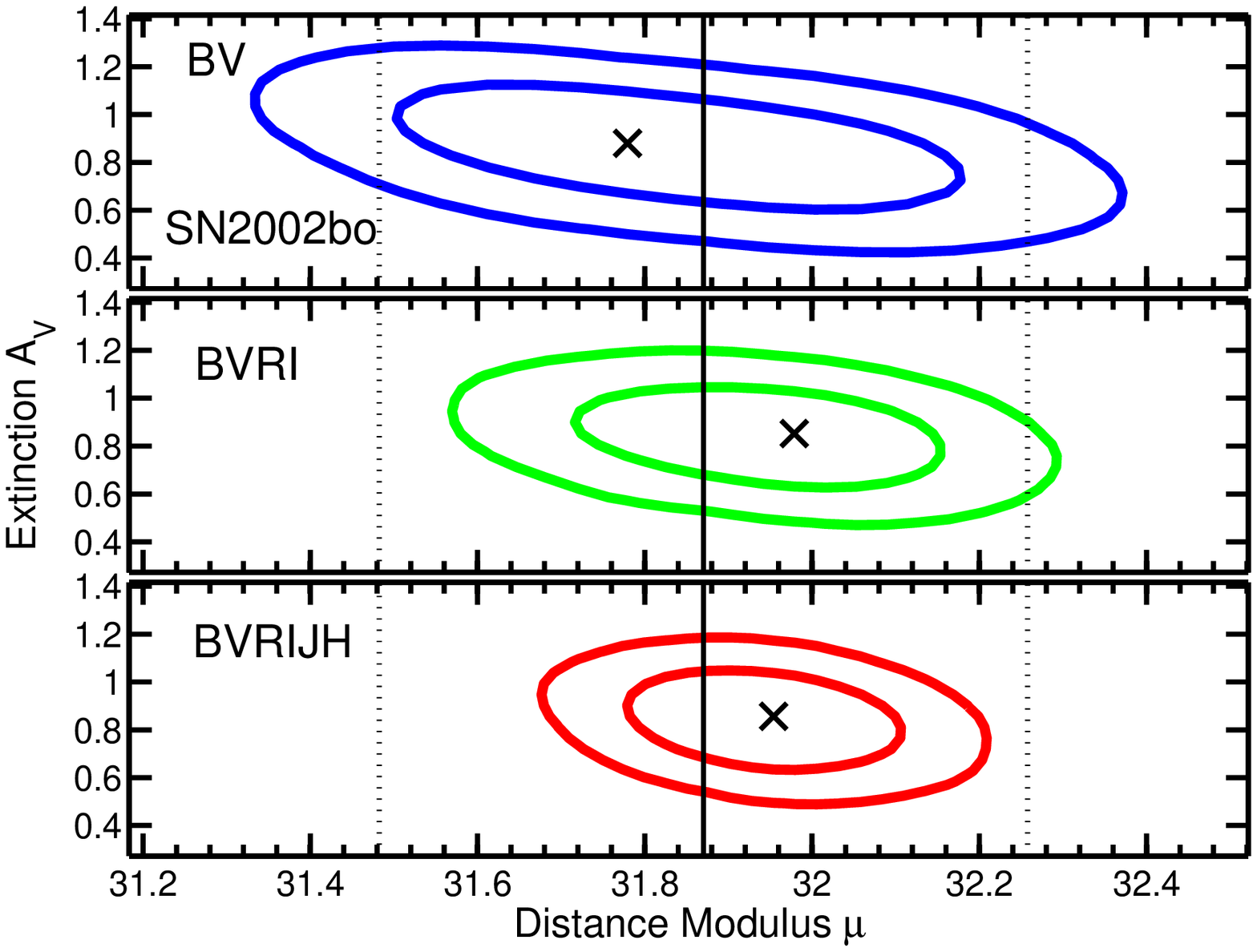}
\caption{\label{fig:omit1} The effect of adding NIR light curve data on statistical uncertainties on distance modulus $\mu$ and extinction $A_V$ for SN 2002bo.  (left)  The predictive probability density of $\mu$ using $BV$ light curve data only (blue), $BVRI$ data only (green), and $BVRIJH$ data (red), computed from the trained optical-near infrared statistical model.  The pdfs integrate to one, so that more precise predictions are taller, and less precise predictions are broader.  The black vertical lines indicate the expected value of $\mu$ given the redshift and the associated error of $\pm 300$ km/s peculiar velocity dispersion.  (right)  The predictive joint probability density of $(\mu, A_V)$ using optical or optical and near infrared data.  The two-dimensional modes are marked, and the inner and outer contours contain 68\% and 95\% of the highest probability regions.  Whereas with $BV$ data only, the distance is uncertain due to the uncertain extinction by host galaxy dust, with BVRIJH data, the uncertainties in extinction, and thus in distance, are reduced significantly.}
\end{figure}

In this section, we demonstrate the effect of using NIR light curve observations in conjunction with optical data for constraining extinction and for making more precise predictions.    In Figure \ref{fig:omit1} we show the posterior predictive densities for the distance modulus and the joint probability densities for SN 2002bo, an event with high extinction.  With the trained probability model, we computed the joint probability $P( \tilde{\mu}, \tilde{A}_V | \, \tilde{\mathcal{D}}_s, \tilde{z}_s; \mathcal{D}, \mathcal{Z})$ under \emph{prediction} where the light curve data $\tilde{\mathcal{D}}_s$ alternately included the SN 2002bo observations in the $BV$, $BVRI$, or $BVRIJH$ filters.  Recall that, under prediction, the tilded redshift $\tilde{z}_s$ is only used for $K$-corrections and Milky Way extinction, but not in the redshift-distance likelihood function.  The dataset  used for training is denoted $\mathcal{D}, \mathcal{Z}$.  We also compute the marginal posterior predictive probability $P( \tilde{\mu} | \, \tilde{\mathcal{D}}_s, \tilde{z}_s; \mathcal{D}, \mathcal{Z})$ for each case.   The probability density in ($\tilde{\mu}, \tilde{A}_V$) integrates over the uncertainties in the SN light curve fit, and the dust and SN populations.  The marginal density in $\tilde{\mu}$ additionally integrates over the uncertainty in $\tilde{A}_V$, as a ``nuisance'' parameter.  These marginal probability densities were computed directly from the MCMC samples, obtained under prediction, using kernel density estimation.  They are not Gaussian approximations of the posterior probability density.

For comparison we mark the expected distance modulus for the observed redshift, and its expected magnitude uncertainty for a 300 km s$^{-1}$ velocity dispersion.  The marginal probability density in $\mu$ integrates to one, so that the taller pdfs make the most precise distance predictions, and the shorter pdfs make the most uncertain predictions.  The figure demonstrates that adding observations in the redward filters greatly improves the precision of distance predictions, with the full optical and NIR data set yielding the greatest precision with this model.   We also show the mode and 68\% and 95\% highest probability density regions of the joint probability $P( \tilde{\mu}, \tilde{A}_V | \, \tilde{\mathcal{D}}_s, \tilde{z}_s; \mathcal{D}, \mathcal{Z})$.  When only the blue $BV$ data is used, there is a strong degeneracy between the uncertainty in distance modulus and the uncertainty in host galaxy dust extinction.   Both the uncertainty in the distance modulus and in dust extinction are reduced when we condition on the available NIR light curve observations.  The predictive precision (the inverse variance) of the distance modulus of an individual SN is, on average, improved by a factor of 2.2 using $BVRI$ and by a factor of 3.6 using $BVRIJH$ data, compared to using $BV$ light curves alone.  The predictive precision is improved by 60\%, on average, using optical and NIR $BVRIJH$ data versus optical $BVRI$ data alone, and can be improved by up to a factor of 2.6, based on the current sample.

\begin{figure}[b]
\centering
\includegraphics[angle=0,scale=0.225]{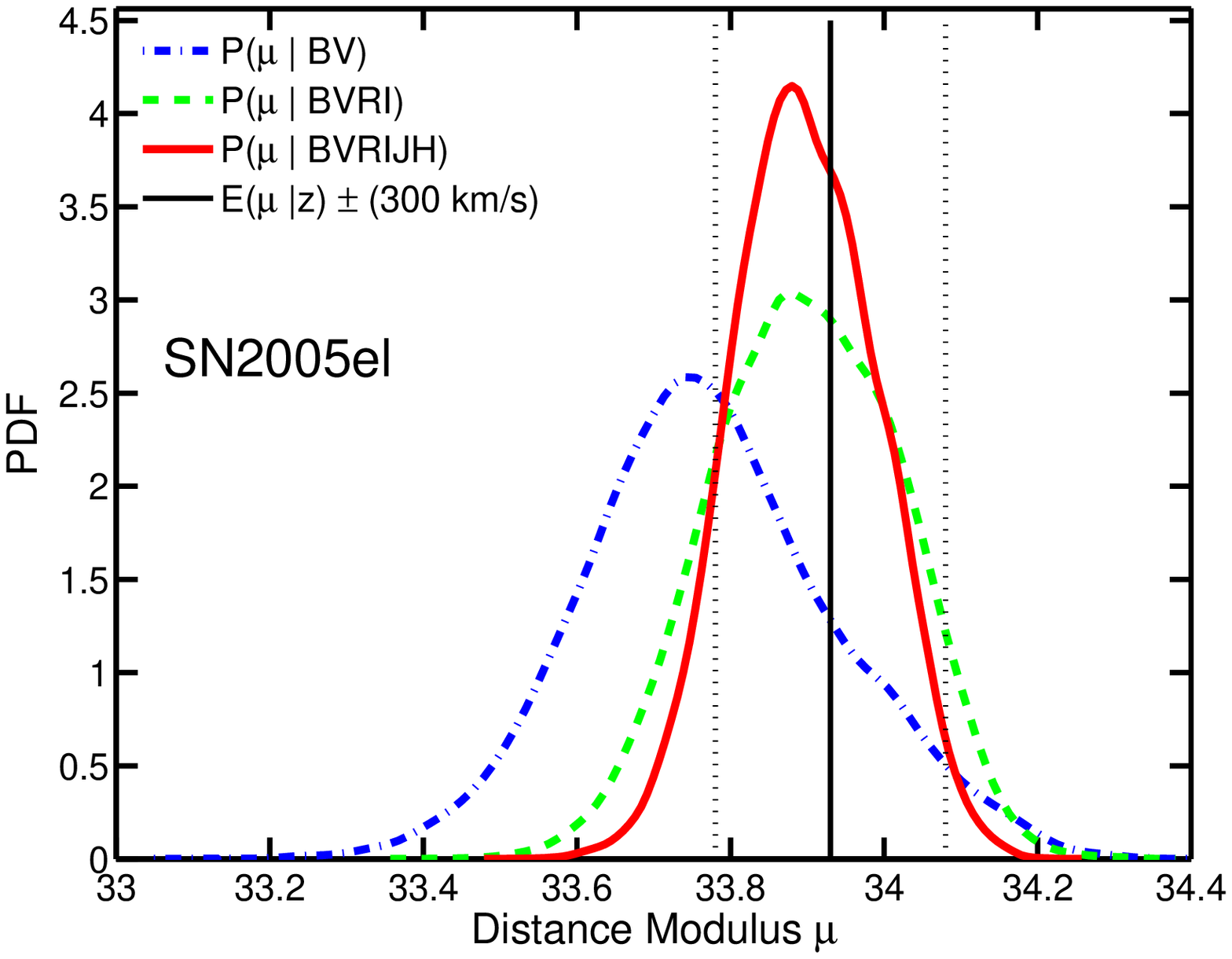}
\hfill
\includegraphics[angle=0,scale=0.225]{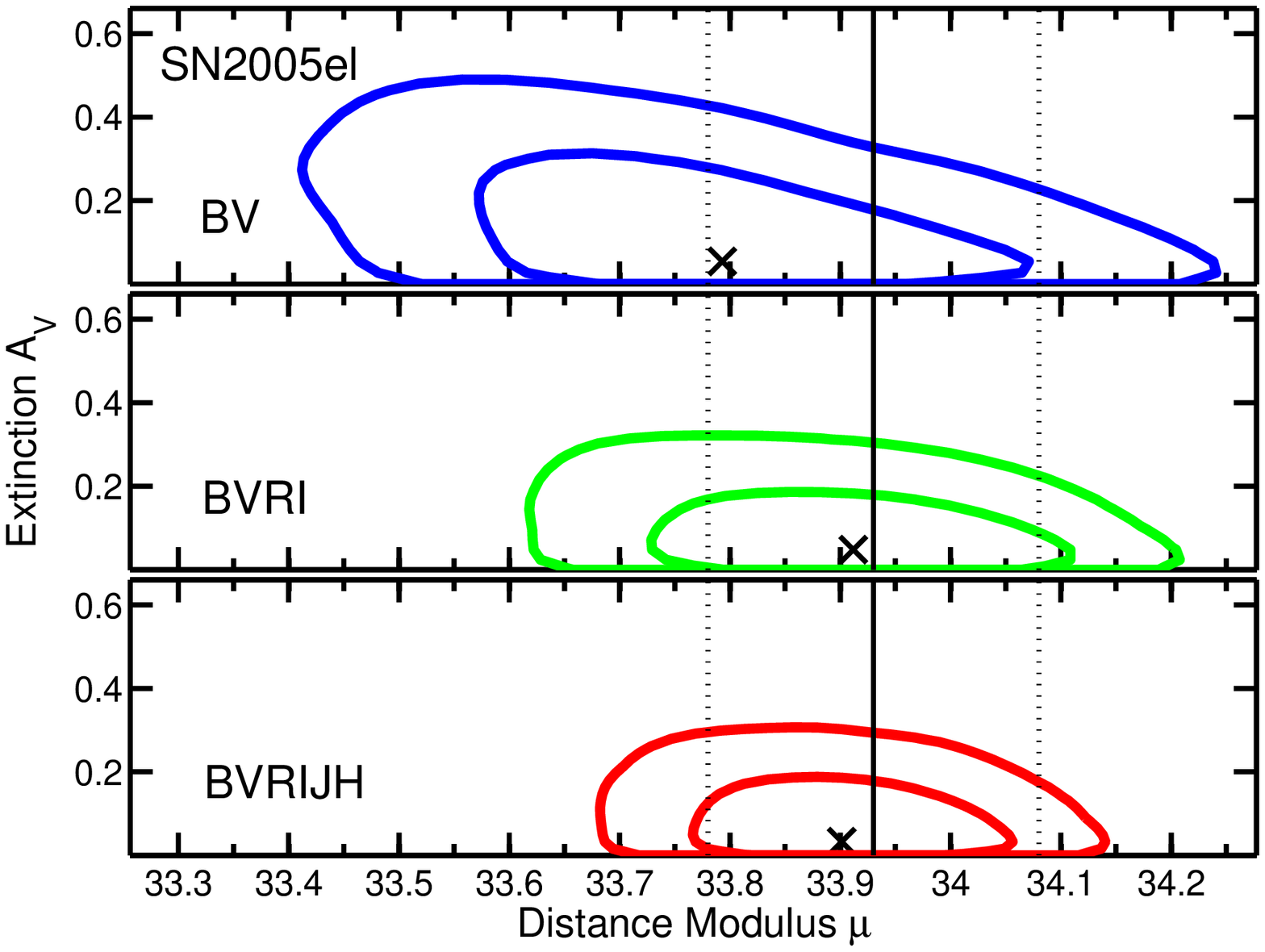}
\caption{\label{fig:omit2}The effect of adding NIR light curve data on statistical uncertainties on distance modulus $\mu$ and extinction $A_V$ for SN 2005el.  (left)  The predictive probability density of $\mu$ using $BV$ light curve data only (blue), $BVRI$ data only (green), and $BVRIJH$ data (red), computed from the trained optical-near infrared statistical model.  The pdfs integrate to one, so that more precise predictions are taller, and less precise predictions are broader.  The black vertical lines indicate the expectation value of $\mu$ given the redshift and the associated error of $\pm 300$ km/s peculiar velocity dispersion.  (right)  The predictive joint probability density of $(\mu, A_V)$ using optical or optical and near infrared data.  The two-dimensional modes are marked, and the inner and outer contours contain 68\% and 95\% of the highest probability regions.  Although this SN most likely has little extinction, the addition of the NIR data still helps to improve the constraints on $A_V$ and the precision of distance predictions.}
\end{figure}

In Figure \ref{fig:omit2}, we illustrate these inferences for a different event, SN 2005el.    This supernova appears to have near zero host galaxy extinction.   However, even with near zero dust extinction there is uncertainty in $A_V$ due to the intrinsic variance of supernova colors.  This uncertainty is in the direction of positive extinction, and hence the joint distribution $P( \tilde{\mu}, \tilde{A}_V | \, \tilde{\mathcal{D}}_s, \tilde{z}_s; \mathcal{D}, \mathcal{Z})$ appears non-Gaussian.
 Although this event has close to zero extinction, there is still a strong degeneracy in the uncertainties  between $\mu$ and $A_V$ under prediction with the $BV$ data alone.    The combination of optical and near infrared data constrains this joint uncertainty and yields improved precision of distance predictions even for low extinction events.
 
\section{Discussion \& Conclusion}\label{discussion}

We have constructed a comprehensive hierarchical model for Type Ia SN light curves in the optical and near infrared ($BVRIJH$).    We model the apparent light curves as the sum of random draws from an absolute light curve population distribution and from a host galaxy population distribution, plus the distance moduli.  While fitting the individual SN Ia light curves, we also estimate the characteristics of the two populations.  These include the intrinsic correlation structure of the absolute light curves, and the joint distribution of extinction $A_V$ and the slope of the dust law $R_V$ in SN Ia host galaxies.   The application of our new \textsc{BayeSN} MCMC algorithm enables coherent probabilistic inference of the unknown parameters and hyperparameters given the observed data.   We also use it to generate distance predictions for SN while marginalizing over the uncertainties in the population models and training set.

The inferred correlation matrices of the intrinsic light curve properties (\S \ref{sec:intrmatrices}) show that the peak optical absolute magnitudes ($BVRI$) are strongly correlated with each other, but have weaker correlation with the $J$ and $H$ near infrared absolute magnitudes.  Similarly, while the peak optical absolute magnitudes are correlated with optical decline rates (particularly $\dmB$), they have low correlation with the NIR decline rates.  The near infrared absolute magnitudes exhibit low correlation with the optical decline rates.  This indicates that the NIR light curves provide independent information on the luminosities of SN Ia, which can be leveraged to improve the precision of distance estimates.

We inferred the distribution of host galaxy extinction $A_V$, with an average value of $\tau_A = 0.37 \pm 0.04$ for our nearby sample.  However, the exponential dust distribution does not adequately fit the fat tail of the peak apparent color distribution: excluding the 4\% of SN in the extreme red tail, the apparent colors and dust extinctions of the other 96\% of SN are well described by an exponential distribution in $A_V$ with $\tau_A = 0.28\pm 0.04$ and an intrinsic color distribution.  Using both linear and step function models (\S \ref{sec:dustpop}), we modeled and inferred the joint distribution of $A_V$ and the extinction law slope parameter, $R_V$, and found strong evidence for an apparent correlation.    Under the assumption of a linear trend between $R_V^{-1}$ and $A_V$, we found a positive slope.  In the limit of low extinction, the marginal estimate of $R_V \approx 2.8 \pm 0.5$ is consistent with the Milky Way interstellar average, $R_V = 3.1$, and with independent measurements of dust in external galaxies \citep[][$R_V = 2.8$]{finkelman08,finkelman10}.
For  SN with very high extinction $A_V > 1$, values of $R_V \approx 1.7 $ are favored.    However, we do not know if the linear assumption is valid over the whole range of $A_V$, so we have explored alternative models for the potential differential behavior of $R_V$.  Under the assumption of a ``step'' function model that groups together SN in four bins in $A_V$, the characteristic $R_V$ in the lowest extinction bin ($A_V < 0.4$)  had a modal value of 2.5 and mean values 2.7-2.9 and  for the highest extinction events ($A_V > 1.25$), we found $R_V = 1.6 \pm 0.1$.   We find that these differences are statistically significant.  

These results suggests that SN at low extinction are seen through lines of sight with ``normal'' interstellar dust, but SN at high extinctions are seen through dust with a steeper reddening law.   This may indicate a circumstellar dust component dominating the absorption of light to high extinction events.    \citet{lwang05} and \citet{goobar08} suggested that scattering of SN light by circumstellar dust clouds could lead to low values of $R_V$.  \citet{goobar08} calculated that multiple scattering of SN light by dust in the locality of the SN would attenuate short wavelength photons and steepen the extinction law to $R_V \sim 1.5-2.5$.  \citet{patat07} reported the detection of spectroscopic signatures of circumstellar material around SN 2006X, a highly extinguished SN Ia.   The effects of circumstellar dust might provide an explanation for the unusual colors of some high extinction events.

With consideration to the uncertainties of our inferences and model assumptions, a conservative conclusion is that most SN in our sample are affected by host galaxy dust with $R_V$ in the range of 2 to 3.  These SN are extinguished by $A_V \lesssim 1$ mag.  At higher extinctions, $A_V \gtrsim 1$, the SN are obscured by dust with $R_V$ in the range of 1.5 to 2, although it is also possible that those SN have different intrinsic colors than the general population.  Notably, we do not find $R_V$ values lower than 1.5.
This is at odds with the $R_V$ values between 1 and 2, many of which were below 1.5,  fit for their whole sample by \citet{folatelli10} by minimizing the scatter in the Hubble diagram of CSP supernovae.  It is also at variance with similarly low $R_V$ values found in the recent literature.  However, those analyses assume that a single $R_V$ value applies to all the SN for a given fit, while we have allowed for distributions in $R_V^{-1}$ that may be dependent on $A_V$.  If we assume that every SN in our sample has exactly the same $R_V$ (complete pooling), or that the distribution of $R_V^{-1}$ has a single mean independent of $A_V$ ($m=0$), we also find $R_V = 1.6-1.7$.  This suggests that these assumptions lead to estimates of $R_V$ that are biased towards smaller values.  This is not surprising, since $R_V$ is best determined for SN with high $A_V$, and as these SN also have apparently low $R_V$, they dominate the estimate of a global constant or average.

Differences are also likely to arise from the treatment of the intrinsic covariance structure of the SN Ia light curves.  In this paper, we have modeled the intrinsic covariances between the absolute light curves in optical and NIR wavelengths spanning $-12$ to $45$ days in phase, and estimated them by probabilistically de-convolving the apparent distributions using a hierarchical model.  Posterior estimates of $A_V, R_V$ for each SN and all other parameters were obtained via Eq. \ref{eqn:globalposterior}, and by marginalizing over uncertainties.  In particular, as estimates of $R_V$ for individual SN at low extinction are difficult to determine exactly, it is necessary to marginalize over that uncertainty when making inferences and predictions, as we do as part of the Bayesian inference.   The inclusion of the NIR light curve data yields an added benefit.  Since the peak $H$-band absolute magnitude is a good standard candle by itself, it is used to predict precise distances that are less susceptible to error from the dust estimate or the dust model.  For example, for a single SN with average extinction, $A_V \approx 0.3$ mag, the change in $A_H$ between $R_V = 1.7$ and $R_V = 3.1$ is about 0.02 mag.

The linear and step function models of the joint distribution of ($A_V, r_V$) both suggest that the average effective $R_V$ at a given level of $A_V$ decreases gradually with the increasing host galaxy extinction.  We might speculate on the existence of two kinds of host galaxy dust with two different reddening laws over wavelength.  One would correspond to ``normal'' interstellar dust as found in the Milky Way, $R_V \approx 3$, and the other would correspond to some kind of circumstellar dust with a reddening law with $R_V \approx 1.7$.  If the dust affecting each SN is comprised of random amounts of these two types of dust, then the effective $r_V$ would roughly be an extinction-weighted average of the characteristic $r_V$s of their respective reddening profiles.  If the circumstellar component was associated with highly dusty environments, then this mixture could generate an apparent trend of effective $r_V$ against total extinction.   This suggests an extension of our hierarchical model, which we will address in a future work.

Using bootstrap cross-validation, we have randomized the optical and near infrared training set to generate probabilistic estimates of the distance moduli to out-of-sample SN.   Comparing these to the distances expected from the Hubble expansion, we found a total \emph{rms} prediction error of 0.16 mag  (at $cz > 3000 \text{ km s}^{-1}$) for SN with optical light curves ($BVRI$) only, but a total \emph{rms} error of  0.11 mag for SN with optical and near infrared ($BVRIJH$) light curves.   After accounting for the dispersion expected from random peculiar velocities with  $\sigma_\text{pec} = (150, 300) \text{ km s}^{-1}$  the \emph{rms} intrinsic prediction errors for these subsets were $(0.15 \pm 0.01, 0.13 \pm 0.02)$ mag for optical and $(0.10 \pm 0.02, 0.08 \pm 0.03)$ mag for optical and NIR.  This demonstrates that distances to SN Ia observed in the optical and near infrared can be estimated with about twice the accuracy ($\sim [0.15/0.10]^2$) of SN Ia observed in the optical alone.  By conditioning on light curve data subsets ($BV$, $BVRI$, $BVRIJH$) for individual SN, we show that including near infrared light curve data tightens the constraints on host galaxy extinction and distance predictions (\S \ref{sec:improving}).

The number of published optical and near infrared light curves of SN Ia  is still small compared to the sample of optically observed events.
Future, larger samples of SN Ia with accurate, joint optical and near-infrared photometry will help test and build the statistical strength of our conclusions on the utility of combining optical and NIR light curves for improving distance predictions and will help illuminate the nature of the dust in SN Ia host galaxies.  In addition to estimating the intrinsic correlation structure of SN Ia light curves and the distribution of host galaxy dust, our hierarchical framework can be applied to distance prediction and analysis of a cosmological sample of SN Ia.  Cosmological samples of SN Ia observed in the rest-frame NIR are possible.  The improved precision and accuracy of the inferences about the history of cosmic expansion may justify the extra effort required to obtain these data now with the \emph{Hubble Space Telescope}, soon with the \emph{James Webb Space Telescope}, and eventually with the \emph{WFIRST} mission.

 \acknowledgements
 
K.M. thanks  Stephan\'e Blondin, Peter Challis, Jonathan Chang, Ryan Foley, Jonathan Foster, Andrew Friedman, Andrew Gelman, Malcolm Hicken, Joseph Koo, Sam Kou and Michael Wood-Vasey for useful discussions, suggestions and clarifications.    We thank the anonymous referee for a thorough review of the manuscript.  Computations in this paper were run on the Odyssey cluster supported by the FAS Sciences Division Research Computing Group at Harvard University.
Supernova research at Harvard College Observatory is supported in part by NSF grant AST-0907903. 

\appendix

\section{Differential Decline Rates Light Curve Model}\label{appendix:ddr}

The continuous normalized light curve is equivalent to the specification of the \emph{total} decline rates of the apparent light curve.   Let $D^F(t)$ be the total decline rate from phase zero to phase $t$: $D^F(t) \equiv \text{LC}^F(t) - F_0 = l^F(t)$.  For example the well-known $B$-band decline rate from peak to 15d past maximum is $\Delta m_{15}(B) = D^B(15)$ \citep{phillips93}.

For practical purposes, it is necessary to parameterize the continuous model for the light curve with a discrete set of variables. Let $\bm{\tau}$ be a grid in rest-frame phase.  The total decline rates to each discrete grid point, $\bm{D}^F = \{D^F_j \equiv D^F(\tau_j)\}$, defines the normalized $F$-band light curve $l^F(t)$ at all times if we choose a suitable interpolation rule.  If we choose a natural cubic spline we ensure continuity up to two derivatives, and the normalized light curve is then linear in the total decline rates: $l^F(t) = \bm{s}(t, \bm{\tau}) \cdot \bm{D}^F$ and the linear smoother $\bm{s}(t, \bm{\tau})$ is specified.  The differential decline rate is $d^F_j \equiv D^F_j - D^F_{j-1}$.    The total decline rates at the knots of $\bm{\tau}$ are sums of the differential decline rates over the span in phase.
There is a simple constant matrix $\bm{G}$ so that $\bm{D}^F = \bm{G} \bm{d}^F$, where $\bm{d}^F$ is the vector of differential decline rates $d^F_j$.  The model for the normalized light curve in band $F$ is linear in the differential decline rates: $l^F(t) = \bm{s}(t, \bm{\tau}) \cdot \bm{G} \bm{d}^F$.   

This Differential Decline Rates representation is a special case of Eq. \ref{eqn:genrep}, with $l^F_0(t) \equiv 0$, $\bm{l}^F_1(t) \equiv \bm{s}(t, \bm{\tau})  \cdot \bm{G}$, $\bm{\theta}^F = \bm{\theta}^F_L \equiv \bm{d}^F$, and $\bm{\theta}_{NL}^F = \emptyset$.
We use an irrregular grid $\bm{\tau}$ in phase spanning $-10$d to $45$d.   The knots are placed more densely near phases where we expect the most observations (near phase zero) and where we expect  more curvature of light curves in certain bands.

\section{$K$-corrections and Milky Way Extinction}\label{appendix:kcorr}

The observed spectral energy distribution (SED) of a SN Ia changes relative to a
fixed observer frame passband due to the effect of cosmological redshift and
extinction in the Milky Way varying with line of sight. To account for these
differences in flux, we derive $K$-corrections and Galactic extinctions for
type Ia supernovae in observer-frame optical and NIR filters. As a high-quality
SED time sequence is seldom available for all supernovae, particularly at
higher redshifts, we use the average spectral template sequence from
\citet{hsiao07}. 

We compute $K$-corrections for optical and NIR passbands following the method
of \citet*{nugent02}. We model the $UBVRI$ filters
with the ``shifted'' Bessell passbands \citep{kessler09} and the $JHK_s$ filters,
with the 2MASS passbands. We choose a standard color that includes each
passband ($U$:$U-B$, $B$:$B-V$, $V$:$V-R$, $R$:$R-I$, $I$:$R-I$, $J$:$J-H$, $H$:$J-H$, $K_s$:$H-K_s$) and for
each model filter, we warp the SED sequence using the $R_V =3.1$ extinction law
\citep{ccm89} to take on a wide range of the corresponding rest-frame color,
measured with synthetic photometry. For a given observer frame passband and
redshift we determine the rest-frame passband with the nearest effective
wavelength. The warped SED series is then redshifted and used to determine the
$K$-correction as a function of rest-frame color. This procedure does not
account for spectral features that vary with lightcurve shape and there is no
constraint on the SED blueward or redward of the bluest or reddest model
filter.

To compute the Milky Way Galactic extinction, we follow a procedure, modified from that outlined in 
\citet*{jha07}.  For a given model filter,
we use the warped SED sequence constructed for the $K$-corrections and
determine the unextinguished observer-frame magnitude with synthetic
photometry as a function of phase and rest-frame color. The sequence is then
reddened with a $R_V = 3.1$ law for a range of values of the Milky Way reddening 
$E_\text{MW} \equiv E(B-V)$ and the
extinguished observer frame magnitude is computed. The difference between the
two magnitudes is the Galactic extinction. We find that the Galactic extinction
for a given phase $t$ and passband $X$ is well modelled by a quadratic in $E_\text{MW}$: $
 A_X(t) = [\alpha_X(t,c) + \beta_X(t,c) E_\text{MW} ] E_\text{MW}$.
We solve for the polynomial co-efficients $\alpha_X$ and $\beta_X$ for all the
phases $t$ and rest-frame apparent color $c$. Further, as the rest-frame color dependence was
introduced by warping the same spectral sequence, we find the co-efficients
$\alpha_X$ and $\beta_X$ to be smoothly varying functions of rest-frame apparent color $c$. We
model the slope and intercept by polynomials of rest-frame color with degree
$4$ and $5$, respectively. We can thereby reduce the Galactic extinction to a
simple set of polynomial coefficients of color tabulated with phase.  The value of 
$E_\text{MW}$ for each SN is obtained from the \citet{sfd98} maps.

\section{Specification of the Hyperprior}\label{appendix:hyperpriors}

There are two populations in this hierarchical model:  the multi-band light curve distribution, and the host galaxy dust extinction distribution for $A_V$ and $R_V$.  The hyperparameters of the SN Ia light curves are population mean $\bm{\mu}_\psi$ and the covariances $\bm{\Sigma}_\psi$.  The hyperparameters of the dust populations are $\tau_A$, $\bm{\beta}$ and $\sigma_r^2$ (e.g. for Case 5 in \S \ref{sec:dustpop}).   We must make explicit our priors on these hyperparameters, i.e. hyperpriors.  At the highest level of the hierarchical model, we use diffuse, or ``non-informative'' prior distributions by default.

The  dust population hyperprior is $P(\tau_A, \bm{\beta}, \sigma_r^2) = P(\tau_A) P(\bm{\beta} | \, \sigma_r^2) P(\sigma_r^2)$.  We adopt uniform prior $P(\bm{\beta} | \sigma_r^2) \propto 1$.   For  $\tau_A$ and $\sigma^2_r$,
we use the standard non-informative prior for positive scale parameters, $P(\log \tau_A ) \propto 1$, $P(\log \sigma_r^2) \propto 1$.

The hyperprior on the absolute light curve distribution hyperparameters  can be conditionally decomposed:
$P(\bm{\mu}_\psi ,\bm{\Sigma}_\psi) = P(\bm{\mu}_\psi | \bm{\Sigma}_\psi) P(\bm{\Sigma}_\psi)$.
We assume a uniform $P(\bm{\mu}_\psi | \, \bm{\Sigma}_\psi ) \propto 1$.  
For $P(\bm{\Sigma}_\psi)$, we require a diffuse density that has support only on the space of symmetric, positive definite, and invertible matrices.   We employ the standard inverse Wishart distribution, which is conjugate to the normal covariance matrix:
$P( \bm{\Sigma}_\psi ) =  \text{Inv-Wishart}_{\nu_0}(\bm{\Sigma}_\psi  | \,\bm{\Lambda_0}) \times f(\bm{\sigma}_\psi)$.
The inverse Wishart density is multiplied by a smooth density on the variances $ f(\bm{\sigma}_\psi)$ described below.  The degrees of freedom parameter is set to $\nu_0 = K +1$ where $K \equiv \text{dim}(\bm{\psi})$.  This guarantees that the marginal prior density of any individual correlation $\rho(\psi_i, \psi_j) \equiv R_\psi^{ij}$ is uniform between $-1$ and $1$ \citep*{barnard00}.  The scale matrix is set to $\bm{\Lambda}_0 = \epsilon_0 \bm{I}$, where $\bm{I}$ is a $K \times K$ identity matrix.   The scale $\epsilon_0$ has the effect of setting a floor on each $\sigma_\psi^i$ so that it does not fall below the value of $\epsilon_0 /\sqrt{N_\text{SN}} \approx 0.02$.  Since we do not realistically expect any standard deviation $\sigma_\psi^i$ to be less than a few 0.01 mag, this limit is conservative, and helps to prevent the MCMC chain from getting stuck in a region of parameter space with a near-zero variance, where the covariance matrix may be nearly singular.

We found it  useful to stabilize the estimation of magnitude variances with a power-law density: $\log f(\bm{\sigma}_\psi) = \log f[\sigma(M_B)] = pN_\text{SN} \log \sigma(M_B)$ with $p \approx 0.09$.   We chose the smallest value of $p$ for which the inferences were locally insensitive to its value.  We have checked that the choice of $p$ does not significantly impact the \emph{rms} prediction error.   We also used the scaled inverse Wishart distribution \citep{gelman&hill, omalley08} as an alternative hyperprior and obtained comparable results.

\section{Mathematical Details: \textsc{BayeSN}}\label{appendix:bayesn}

In this section, we provide mathematical results for each step of the \textsc{BayeSN} algorithm.

\begin{enumerate}

\item  The goal is to sample from $P(\bm{\mu}_\psi, \bm{\Sigma}_\psi | \, \cdot, \mathcal{D}, \mathcal{Z} )$.  This can be factored as $P(\bm{\mu}_\psi | \,  \bm{\Sigma}_\psi,  \{\bm{\psi}_s \}) P( \bm{\Sigma}_\psi | \, \{\bm{\psi}_s \})$.    These densities only depend on $\{ \bm{\psi}_s \}$ through the sufficient statistics:  the sample mean $\bar{\bm{\psi}}$, and the matrix sum of squared deviations from the mean: $\bm{S}_\psi= \sum_{s=1}^{N_\text{SN}} ( \bm{\psi}_s -  \bar{\bm{\psi}} )( \bm{\psi}_s -  \bar{\bm{\psi}} )^T$.
We generate a new $\bm{\Sigma}_\psi^*$ from the proposal density
$q(\bm{\Sigma}_\psi^* | \, \{ \bm{\psi}_s \} ) = \text{Inv-Wishart}_{\nu_N}(\bm{\Sigma}_\psi^*| \, \bm{\Lambda}_N^{-1})$,
where $\nu_N = \nu_0 +N_\text{SN} $, and $\bm{\Lambda}_N = \bm{S}_\psi + \bm{\Lambda_0}$.   When the $(\epsilon_0 / N_\text{SN} )^2$ is negligible compared to the variances, the expectation of this distribution is just the standard maximum likelihood estimator of covariance, $\bm{S}_\psi / N_\text{SN}$.
If $f(\bm{\sigma}_\psi) \propto 1$ then the proposal is the same as $P(\bm{\Sigma}_\psi^* | \, \{ \bm{\psi}_s \})$, and this is Gibbs sampling.  If not, then the Metropolis-Hastings ratio $r$ simplifies to $r = f(\bm{\sigma}_\psi^*) / f(\bm{\sigma}_\psi)$.  The proposal is accepted ($\bm{\Sigma}_\psi \rightarrow \bm{\Sigma}_\psi^* $) with probability $r$.  This method results in fast convergence since it allows for updating the entire covariance matrix at once.
A new $\bm{\mu}_\psi$ is Gibbs sampled from $P(\bm{\mu}_\psi | \,  \bm{\Sigma}_\psi,  \{\bm{\psi}_s \}) = N(\bm{\mu}_\psi | \, \bar{\bm{\psi}}, \bm{\Sigma}_\psi / N_\text{SN})$.

\item  The conditional density for $\tau_A$ is $P( \tau_A | \, \cdot, \mathcal{D}, \mathcal{Z} ) = P( \tau_A | \, \{A_V^s\}) = \text{Inv-Gamma}(\tau_A | \, \nsn, \sum_{s=1}^\nsn A_V^s)$.

\item  The conditional densities $P(\bm{\beta} | \, \sigma^2_r, \{r_V^s, A_V^s\} )$ and $P(\sigma^2_r | \, \{r_V^s, A_V^s\})$ are standard results of Bayesian analysis of ordinary linear regression of $r_V$ versus $A_V$ \citep[][Ch. 14]{gelman_bda}.

\item Since the subsequent steps concern only one SN at a time, we suppress the label $s$ on individual SN parameters.

\begin{enumerate}

\item  The conditional posterior density of the fit $(T_0, \bm{\phi})$ for a single SN is proportional to
\begin{equation}\label{eqn:4a1}
P(T_0, \bm{\phi} | \, \cdot, \mathcal{D}_s, z_s) \propto P( \bm{m} |\, T_0, \bm{\phi}, z_s) \times N(\bm{\phi} | \, \bm{\mu}_\phi, \bm{\Sigma}_\psi)
\end{equation}
where $\bm{\mu}_\phi \equiv  \bm{\mu}_\psi + \bm{A} + \bm{v}\mu$, and the first factor is Eq. \ref{eqn:lc_lkhd}.  We construct a proposal density for the new fit $(T_0^*, \bm{\phi}^*)$ given the current one:  $q(T_0^*, \bm{\phi}^* | \, T_0, \bm{\phi}) = q(\bm{\phi}^* | \, T_0^*; T_0, \bm{\phi}) \times q( T_0^* | \, T_0;  \bm{\phi})$.
The proposal for the new $T_0^*$ is $q( T_0^* | \, T_0;  \bm{\phi}) = N( T_0^* | \, T_0 , s^2_T)$.   The proposal $q(\bm{\phi}^* | \, T_0^*; T_0, \bm{\phi})$ is an approximation to Eq. \ref{eqn:4a1} with the $K$-correction and Milky Way extinction factors fixed at the current fit $(T_0, \bm{\phi})$.   These depend on $\bm{\phi}$ only through the apparent colors.  The proposal is 
\begin{equation}\label{eqn:4a2}
q(\bm{\phi}^* | \, T_0^*; T_0, \bm{\phi}) \propto N[\bm{m} | \, \text{\bf KC}(T_0; z, \bm{\phi})+ \text{\bf GX}(T_0; z, \bm{\phi}, E_{\text{MW}}) + \bm{L}_2(T_0^*, z) \bm{\phi}^*, \bm{W} ] \times N(\bm{\phi}^* | \, \bm{\mu}_\phi, \bm{\Sigma}_\psi)
\end{equation}
After algebraic simplifications, it can be shown that this is a Gaussian probability density on $\bm{\phi}^*$ and thus can be used to generate a random proposal.  The joint proposal $(T_0^*, \bm{\phi}^*)$ is accepted with probability
\begin{equation}
r = \frac{P(T_0^*, \bm{\phi}^* | \, \cdot, \mathcal{D}_s, z_s)}{P(T_0, \bm{\phi} | \, \cdot, \mathcal{D}_s, z_s)} \times  \frac{q(\bm{\phi} | \, T_0;  T_0^*, \bm{\phi}^*)}{q( \bm{\phi}^* | \, T_0^*; T_0, \bm{\phi})}.
\end{equation}
The rejection step corrects the approximation of the conditional, Eq. \ref{eqn:4a1}, with the proposal Eq. \ref{eqn:4a2}.  If the $\text{\bf KC}$ and $\text{\bf GX}$ factors are constant with respect to SN color and phase, then $r=1$ and this is just Gibbs sampling.  This scheme is efficient when $\text{\bf KC}$ and $\text{\bf GX}$ are slowly varying with phase and apparent color.

\item The conditional density for $\mu$ simplifies to  $P(\mu | \, \cdot, \mathcal{D}_s, z_s) = N( \mu | \, \hat{\mu}, \hat{\sigma}_\mu^2)$, where
$\tilde{\mu} = s_\mu^2 \bm{v}^T \bm{\Sigma}_\psi^{-1} ( \bm{\phi} - \bm{A} - \bm{\mu}_\psi)$;  $s_\mu^{-2} = \bm{v}^T \bm{\Sigma}_\psi^{-1} \bm{v}$; 
 $\hat{\sigma}_\mu^{-2} = s_\mu^{-2} + \sigma_\mu^{-2}$; and $\hat{\mu} = \hat{\sigma}_\mu^{2} ( \sigma_\mu^{-2} f(z) + s_\mu^{-2} \tilde{\mu})$. For prediction, we take $\sigma_\mu \rightarrow \infty$.

\item   The conditional  density for $A_V$ is $P(A_V | \, \cdot, \mathcal{D}_s, z_s) = P(A_V | \bm{\phi}, \mu, r_V; \bm{\mu}_\psi, \bm{\Sigma}_\psi, \tau_A, \bm{\beta}, \sigma^2_r)$.  This is a probability density on $A_V \ge 0$ proportional to $N(A_V | \hat{A}, s_A^2) \times N( r_V | \, \beta_0 + \beta_1 A_V, \sigma^2_r)$, where $\hat{A} = s_A^2  \bm{c}^T \bm{\Sigma}_\psi^{-1} [ \bm{\phi}  - \bm{v} \mu - \bm{\mu}_\psi] -s_A^2 /\tau_A$; $\bm{c} = (\bm{\alpha} + \bm{\beta} r_V)$, and $ s_A^{-2} = \bm{c}^T \bm{\Sigma}_\psi^{-1} \bm{c}$.  This can be sampled using griddy Gibbs sampling.

\item   The conditional posterior $P(r_V | \, \cdot, \mathcal{D}_s, z_s) = P(r_V |  \bm{\phi}, \mu, A_V;  \bm{\mu}_\psi, \bm{\Sigma}_\psi, \bm{\beta}, \sigma^2_r)$.  Defining $\tilde{\sigma}_r^{-2} = A_V^2 \bm{\beta}^T \bm{\Sigma}_\psi \bm{\beta}$; 
$\tilde{\mu}_r = \tilde{\sigma}^2_r   \bm{\beta}^T A_V \bm{\Sigma}_\psi^{-1} [ \bm{\phi}  -\bm{v} \mu - A_V \bm{\alpha} - \bm{\mu}_\psi]$;
$\hat{\sigma}_r^{-2} = \tilde{\sigma}^{-2}_r + \sigma^{-2}_r $; $\hat{r}_V = \hat{\sigma}^2_r  [ \tilde{\sigma}^{-2}_r \tilde{\mu}_r + \sigma^{-2}_r (\beta_0 + \beta_1 A_V)]$, this density is proportional to $N(r_V | \, \hat{r}_V, \hat{\sigma}_r^2)$ over the restricted range $0.18 < r_V  < 0.7$.  A new sample is generated by evaluating the pdf on a fine grid and using griddy Gibbs sampling.

\item (optional) Generalized conditional sampling allows the MCMC to move along expected degeneracies between parameters in the posterior density that may be oblique with respect to the natural coordinate system defined by the chosen parameters \citep{liu&sabatti00, liu02}.  We expect there to be a trade-off between dust extinction and distance to SN, since both make SN appear dimmer.  Let $p(A_V, \mu) = P(A_V, \mu | \, \cdot, \mathcal{D}_s, z_s)$ be the conditional posterior of dust and distance.  To perform the translation $(A_V, \mu) \rightarrow (A_V, \mu) + \gamma (1, -x)$, we first choose a scalar $x$ which sets a direction in the $(A_V, \mu)$  plane to move along. 
To select an appropriate direction along the trade-off between dust and distance, we find $\bar{x} = \min_x  | (\bm{\alpha} + \bm{\beta} r_V) - x \bm{v} |^2$.
For typical values of $r_V$, this was $\bar{x} \approx 0.7$.   To select a translation vector near this direction, we sample $x \sim N(0.7, 0.05)$.  Then we sample a random $\gamma \sim p(A_V + \gamma, \mu -x \gamma)$, where $A_V$ and $\mu$ are the current values.  The sample can be generated by evaluating the univariate density on a grid and using the inverse cdf method.  Given $\gamma$ the chain can be translated to the new position.

\end{enumerate}
\end{enumerate}

\bibliographystyle{apj}
\bibliography{apj-jour,sn,stat}{}

\clearpage

\input{sn_lc_dust_table_stub2.tex}

\input{sn_mu_table_stub.tex}

 \clearpage

\end{document}

%% file: Rv_scenarios.tex
\begin{deluxetable}{rrrr}
\tabletypesize{\scriptsize}
\tablecaption{$\mu$ Prediction Errors for $R_V$ scenarios\label{table:Rv}}
\tablewidth{0pt}
\tablehead{ \colhead{Assumptions} & \colhead{Inferred} & \colhead{Opt.} & \colhead{Opt+NIR } \\
\colhead{on $R_V$ population} & \colhead{Hyperparameters} & \colhead{[mag]} & \colhead{[mag]} } 
\startdata
$R_V=3.1$ &  \nodata &  0.20 &  0.13 \\ 
 Complete Pooling  & $R_V = 1.6 \pm 0.1$  &  0.15 &  0.13 \\ 
 No Pooling &  \nodata &  0.16 &  0.12 \\ 
 PP: m = 0 &  $\mu_r^{-1} =  1.7 \pm 0.1$, &  0.16 &  0.12 \\ 
  &   $\sigma_r =  0.04 \pm 0.02 $  & & \\
 PP: m = 1 &  $\beta_0 = 0.35 \pm 0.05$  &  0.15 &  0.11 \\ 
  & $\beta_1 = 0.15 \pm 0.03$ & & \\
   &$\sigma_r = 0.04 \pm 0.02$ & & \\
PP: 4-Steps & c.f. Table \ref{table:Rv_4steps} & 0.15 & 0.11
\enddata
 \tablecomments{Optical and Optical+NIR rms prediction errors at $cz > 3000 \text{ km s}^{-1}$ for different dust population models.  Estimates of hyperparameters are the marginal posterior means and standard deviations.  The rms prediction errors are the 0.632 bootstrap cross-validation estimates. Sampling variance of prediction errors is typically $\pm 0.01$ mag.}
\end{deluxetable}

%% file: Rv_steps2.tex
\begin{deluxetable}{rrrrr}
\tabletypesize{\scriptsize}
\tablecaption{ Inference for ``2-Step'' $R_V$ Population Model\label{table:Rv_2steps}}
\tablewidth{0pt}
\tablehead{ \colhead{$A_V$ Range} & \colhead{$\mu_r^{-1}$} & \colhead{$\mu_r$} &\colhead{$\sigma_r$} & \colhead{$p_\text{tail}$}} 
\startdata
$[ 0, 0.8]$ & $2.3 \pm 0.3$ & $0.45 \pm 0.06$ & $0.04 \pm 0.02$ & 0.008 \\ 
$ > 0.8 $ & $1.7 \pm 0.1$ & $0.59 \pm 0.04$ & $0.06 \pm 0.03$ & \nodata
 \enddata
 \tablecomments{The hyperparameter $\mu_r$ is the population mean $R_V^{-1}$ for SN in each interval in $A_V$.  The hyperparameter $\sigma_r^2$ is the population variance of $R_V^{-1}$ in each interval.  Estimates are marginal posterior means and standard deviations.  The marginal posterior density of $\sigma_r$ is highly non-Gaussian.  The estimate of $\mu_r^{-1}$ is not precisely the inverse of the estimate of $\mu_r$, because of uncertainty and the non-linear transformation.  The marginal probability that $\mu_r$ is larger than $\mu_r$ of the high extinction bin is $p_\text{tail}$.}
\end{deluxetable}

%% file: Rv_steps4.tex
\begin{deluxetable}{rrrrr}
\tabletypesize{\scriptsize}
\tablecaption{ Inference for ``4-Step'' $R_V$ Population Model\label{table:Rv_4steps}}
\tablewidth{0pt}
\tablehead{ \colhead{$A_V$ Range} & \colhead{$\mu_r^{-1}$} & \colhead{$\mu_r$} &\colhead{$\sigma_r$} & \colhead{$p_\text{tail}$} } 
\startdata
$[ 0, 0.4]$ & $2.9 \pm 0.7$ & $0.35 \pm 0.08$ & $0.03 \pm 0.02$ & $< 0.001$ \\ 
$[0.4, 0. 8]$  & $2.3 \pm 0.3$  & $0.45 \pm 0.06$ & $0.03 \pm 0.03$ & $0.005$ \\ 
$[0.8, 1.25]$ & $2.1 \pm 0.2$ & $0.48 \pm 0.04$ & $0.03 \pm 0.03$ & $0.004$ \\ 
 $> 1.25$ &  $ 1.6 \pm 0.1$ & $0.63 \pm 0.03$ & $0.04 \pm 0.03$  & \nodata 
 \enddata
 \tablecomments{The hyperparameter $\mu_r$ is the population mean $R_V^{-1}$ for SN in each interval in $A_V$.  The hyperparameter $\sigma_r^2$ is the population variance of $R_V^{-1}$ in each interval.  Estimates are marginal posterior means and standard deviations.  The marginal posterior density of $\sigma_r$ is highly non-Gaussian.  The estimate of $\mu_r^{-1}$ is not precisely the inverse of the estimate of $\mu_r$, because of uncertainty and the non-linear transformation. The marginal probability that the mean $\mu_r$ of each bin is larger than $\mu_r$ of the highest extinction bin is $p_\text{tail}$.}
\end{deluxetable}

%% file: sn_lc_dust_table_stub2.tex
\begin{deluxetable*}{lrrrrrrrrrr}
\tabletypesize{\scriptsize}
\tablecaption{Apparent Light Curve and Dust Estimates for Individual SN Ia\label{table:sn_lcdust}}
\tablewidth{0pt}
\tablehead{ \colhead{SN} & \colhead{$B_0$}\tablenotemark{a} & \colhead{$\Delta m_{15}(B)$} & \colhead{$V_0$}\tablenotemark{b} & \colhead{$R_0$}  & \colhead{$J_0$} & \colhead{$H_0$}  & \colhead{$\hat{A}_V$}\tablenotemark{c} & \colhead{68\%($A_V$)}\tablenotemark{d} & \colhead{$R_V$}\tablenotemark{e} & \colhead{Ref.}\tablenotemark{f} }
\startdata
SN1998bu & $12.11 \pm 0.01$ & $1.03 \pm 0.02$ & $11.80 \pm 0.01$ & $11.65 \pm 0.01$  & $11.74 \pm 0.02$ & $11.87 \pm 0.03$ & 0.97 & [0.85, 1.07] & $2.2 \pm 0.3$ & J99,H00  \\ 
SN1999cl & $14.86 \pm 0.03$ & $1.17 \pm 0.07$ & $13.74 \pm 0.02$ & $13.27 \pm 0.03$  & $12.96 \pm 0.04$ & $13.03 \pm 0.04$ & 1.95 & [1.86, 2.10] & $1.6 \pm 0.1$ & K00  \\ 
SN2005el & $14.85 \pm 0.02$ & $1.28 \pm 0.05$ & $14.91 \pm 0.02$ & $14.97 \pm 0.03$  & $15.53 \pm 0.02$ & $15.75 \pm 0.03$ & 0.01 & [0.00, 0.11] & $2.8 \pm 0.6$ & WC3  \\ 
SN2005eq & $16.26 \pm 0.04$ & $0.88 \pm 0.05$ & $16.23 \pm 0.03$ & $16.33 \pm 0.04$  & $17.01 \pm 0.02$ & $17.26 \pm 0.05$ & 0.25 & [0.16, 0.40] & $2.6 \pm 0.5$ & WC3  \\ 
SN2006ax & $15.01 \pm 0.02$ & $1.05 \pm 0.03$ & $15.08 \pm 0.02$ & $15.18 \pm 0.02$ & $15.87 \pm 0.02$ & $16.29 \pm 0.04$ & 0.01 & [0.00, 0.12] & $2.8 \pm 0.6$ & WC3  \\ 
\enddata
\tablenotetext{a}{Apparent magnitude at maximum light in rest frame $B$ filter, corrected for Milky Way extinction and $K$-corrections.  Estimates only listed if SN was observed in the filter.}
\tablenotetext{b}{Apparent magnitude in rest frame $V$ at time of maximum in $B$, corrected for Milky Way extinction and $K$-corrections.}
\tablenotetext{c}{Marginal posterior mode of extinction $A_V$.}
\tablenotetext{d}{Highest posterior density interval containing 68\% of marginal probability.}
\tablenotetext{e}{Marginal posterior mean and standard deviation.}
\tablenotetext{f}{Reference codes: CfA3: \citet{hicken09a}; WV08: \citet[PAIRITEL; ][]{wood-vasey08}; WC3: WV08+CfA3; J99: \citet{jha99}; H00: \citet{hernandez00}; K00: \citet{krisciunas00}; K01: \citet{krisciunas01}; DP02: \citet{dipaola02}; V03: \citet{valentini03}; K03: \citet{krisciunas03}; K04b: \citet{krisciunas04b}; K04c: \citet{krisciunas04c}; K07: \citet{krisciunas07}; ER06: \citet{elias-rosa06}; ER07: \citet{elias-rosa07}; Pa07: \citet{pastorello07a}; St07: \citet{stanishev07}; P08: \citet{pignata08}.}
\tablecomments{This table is a representative stub. The $I$-band estimate is omitted here to preserve width.}
\end{deluxetable*}

%% file: sn_mu_table_stub.tex
\begin{deluxetable*}{lrrrrrrrr}
\tabletypesize{\scriptsize}
\tablecaption{Distance Modulus Predictions for SN Ia \label{table:mu}}
\tablewidth{0pt}
\tablehead{ \colhead{SN} & \colhead{$cz$} & \colhead{$\mu_{\text{LCDM}}|z$} & \colhead{$\sigma_{\mu}|z$} & \colhead{$\mu_{\text{resub}}$} & \colhead{$\bar{\mu}_{\text{pred}}$} & \colhead{$s_{\text{pred}}$} & \colhead{$\sigma_{\text{pred}}$} & \colhead{$\bar{A}_V^{\text{pred}}$} \\
\colhead{} & \colhead{$[\text{km s}^{-1}]$} & \colhead{[mag]} & \colhead{[mag]} & \colhead{[mag]}  & \colhead{[mag]}  & \colhead{[mag]}  & \colhead{[mag]}  & \colhead{[mag]}  }
\startdata
SN1998bu & 708.90 & 29.97 & 0.46 & 30.00 & 29.95 & 0.02 & 0.10 & 0.97 \\ 
SN1999cl & 957.00 & 30.62 & 0.39 & 30.94 & 30.94 & 0.05 & 0.12 & 1.94 \\ 
SN2005el & 4349.10 & 33.93 & 0.08 & 33.89 & 33.87 & 0.04 & 0.11 & 0.03 \\ 
SN2005eq & 8535.00 & 35.42 & 0.04 & 35.49 & 35.49 & 0.03 & 0.11 & 0.22 \\ 
SN2006ax & 5391.00 & 34.40 & 0.06 & 34.36 & 34.34 & 0.02 & 0.10 & 0.03 \\ 
\enddata
 \tablecomments{This table is a representative stub. $\mu_{\text{LCDM}}|z$ is the distance modulus expected from the redshift assuming $h = 0.72, \Omega_M = 0.27, \Omega_\Lambda = 0.73, w = -1$.   Its magnitude variance, assuming peculiar velocity dispersion $\sigma_{\text{pec}} = 150 \text{ km s}^{-1}$ is  $\sigma^2_\mu$.  $\mu_{\text{resub}}$ is the distance modulus estimated under resubstitution.  Under bootstrap cross-validation, the mean prediction over bootstraps is $\bar{\mu}_{\text{pred}}$ and the standard deviation of predictions over boostraps is $s_{\text{pred}}$.  Zero values of $s_{\text{pred}}$ are less than 0.005 mag.  The average standard deviation of uncertainty of a predictions is $\sigma_{\text{pred}}$.  The marginal posterior mode of $A_V$ under prediction, averaged over the prediction sets for each SN is $\bar{A}_V^{\text{pred}}$.}
\end{deluxetable*}